\documentclass[a4paper,11pt]{article}
\usepackage{jcappub} 
\usepackage{mathtools}
\usepackage{comment}
\usepackage{enumerate}
\usepackage{mathrsfs}
\usepackage{units}
\usepackage{graphicx,amsfonts,amssymb,amsthm,amsmath,psfrag}
\usepackage[toc,page]{appendix}
\usepackage{subfigure}
\usepackage{slashed}
\usepackage{mathtools}
\usepackage{tikz}
\usepackage{booktabs}
\usepackage{siunitx}
\usepackage{bm}         
\usepackage{color}
\usepackage{amssymb}
\usepackage{amsmath}     
\usepackage{psfrag}
\usepackage{graphicx}
\usepackage{epstopdf}
\usepackage{dcolumn}
\usepackage{bm}
\usepackage[utf8]{inputenc}
\usepackage{enumitem}
\usepackage{subfigure}
\usepackage{braket}
\usepackage[color=orange]{todonotes}

\definecolor{asparagus}{rgb}{0.53, 0.66, 0.42}
\hypersetup{linkcolor=blue,filecolor=blue,urlcolor=blue,citecolor=blue}
\usepackage[normalem]{ulem}	
\usepackage{bbding}
\usepackage{amsmath}
\usepackage{multirow}

\newcommand{\cn}{\mathrm{cn}}

\newcommand\redsout{\bgroup\markoverwith{\textcolor{purple}{\rule[0.5ex]{2pt}{0.8pt}}}\ULon}

\title{
\huge{Hubble-induced phase transitions on the lattice with applications to Ricci reheating
}}
\author[a]{Dario Bettoni,}
\emailAdd{bettoni@usal.es}  
\author[b,c]{Asier Lopez-Eiguren,}
\emailAdd{asier.lopez\_eiguren@tufts.edu}
\author[d]{Javier Rubio}
\emailAdd{javier.rubio@tecnico.ulisboa.pt} 
\vspace{5cm}
\affiliation[a]{Departamento de F\'isica Fundamental and IUFFyM, Universidad de Salamanca, \\Plaza de la Merced S/N, E-37008  Salamanca, Spain}
\affiliation[b]{Institute of Cosmology, Department of Physics and Astronomy, Tufts University, \\Medford, MA 02155, USA}
\affiliation[c]{Department of Physics and Helsinki Institute of Physics, \\  PL 64, FI-00014 University of Helsinki, Finland} 
\affiliation[d]{Centro de Astrof\'{\i}sica e Gravita\c c\~ao  - CENTRA,
Departamento de F\'{\i}sica, Instituto Superior T\'ecnico - IST,
Universidade de Lisboa - UL,
Av. Rovisco Pais 1, 1049-001 Lisboa, Portugal}

\abstract{
Using 3+1 classical lattice simulations, we follow the symmetry breaking pattern and subsequent non-linear evolution of a spectator field non-minimally coupled to gravity when the post-inflationary dynamics is given in terms of a stiff equation-of-state parameter. We find that the gradient energy density immediately after the transition represents a non-negligible fraction of the total energy budget, steadily growing to equal the kinetic counterpart. This behaviour is reflected on the evolution of the associated equation-of-state parameter, which approaches a universal value $1/3$, independently of the shape of non-linear interactions.  Combined with kination, this observation allows for the generic onset of radiation domination for arbitrary self-interacting potentials, significantly extending previous results in the literature. The produced spectrum at that time is, however, non-thermal, precluding the naive extraction of thermodynamical quantities like temperature. Potential identifications of the spectator field with the Standard Model Higgs are also discussed.}

\begin{document}
\maketitle

\section{Introduction}\label{sec:intro}

The appearance of tachyonic instabilities is an ubiquitous phenomenon in the physics of the very early Universe, being generically associated with the amplification of quantum fluctuations. Typical examples include natural and new inflation models \cite{Cormier:1998nt,Tsujikawa:1999ni,Rubio:2019ypq,Karam:2021sno}, preheating scenarios with trilinear interactions \cite{Dufaux:2006ee,Abolhasani:2009nb}, spontaneous symmetry breaking settings such as hybrid inflation \cite{Felder:2000hj,Felder:2001kt,Copeland:2002ku,GarciaBellido:2002aj} and second-order phase transitions \cite{Giblin:2011yh}. 

On general grounds, a tachyonic instability emerges whenever a mass eigenvalue in an interacting or self-interacting theory becomes negative. This transition can be caused by a bare mass parameter in the action or effectively induced by the expectation value of a given field or operator. In this context, models in which the instability is triggered by a non-minimal coupling to gravity have recently received considerable attention. 

Hubble-induced phase transitions for spectator fields have been advocated to address several phenomenological aspects such as the generation of the observed matter-antimatter asymmetry \cite{Bettoni:2018utf}, the formation of topological defects with observable gravitational wave backgrounds \cite{Bettoni:2018pbl,Bettoni:2019dcw}, the onset of the hot big bang era \cite{Figueroa:2016dsc,Dimopoulos:2018wfg,Opferkuch:2019zbd} or the non-thermal production of dark matter  \cite{Fairbairn:2018bsw,Laulumaa:2020pqi}. 
The main assumption behind these proposals is the existence of a cosmological expansion era with a stiff equation-of-state parameter ($w > 1/3$) prior to the beginning of radiation domination.~\footnote{For a review of the different expansion histories in the early Universe, see for instance Ref.~\cite{Allahverdi:2020bys}.} This unusual equation of state induces a change of sign in the Ricci scalar and hence in the effective mass for the spectator field, triggering its displacement towards new time-dependent minima appearing at large field values.

In many instances, the evolution of the spectator field during Hubble-induced phase transitions is discussed in terms of an homogeneous component rolling down an effective potential and eventually oscillating around the new symmetry-breaking minima in a fully coherent fashion \cite{Figueroa:2016dsc,Dimopoulos:2018wfg,Opferkuch:2019zbd}.
This naive picture is, however, inaccurate \cite{Felder:2000hj,Felder:2001kt}. On the one hand,  it fails to capture one of the main features of spontaneous symmetry breaking: the production of topological defects \cite{Bettoni:2018pbl}. On the other hand, the dynamics of the system is not dominated by an homogeneous component, but rather by the rapid amplification of long wavelength fluctuations that become relevant well before the field reaches the new minima of the potential \cite{Bettoni:2019dcw}. Among other effects, this has important consequences for the virialization and thermalization of the system. 

In order to verify and assess the relevance of the aforementioned features,  we explore here, for the first time, the non-linear dynamics of Hubble-induced phase transitions using $3+1$ classical lattice simulations. To this end, we focus on a $Z_2$ symmetric spectator field non-minimally coupled to gravity, albeit our results can be easily extended to other global symmetry breaking patterns. 

The present analysis differs in various aspects from previous studies in the context of hybrid inflation and phase transitions, for which lattice simulations are also available (see e.g.  \cite{Felder:2000hj,Felder:2001kt,Copeland:2002ku,GarciaBellido:2002aj,Giblin:2011yh}), and multifield inflation with non-minimal couplings \cite{Nguyen:2019kbm,vandeVis:2020qcp}. First, the tachyonic mass leading to the spontaneous breaking of the aforementioned $Z_2$ symmetry is time-dependent. Second, the system is completely characterized by a single physical scale, the Hubble rate, which controls both the dilution and the tachyonic enhancement of perturbations. When numerical tools are brought to bear, we find also several interesting outcomes: 
\begin{enumerate}[label=\roman*)]
    \item The Hubble-induced tachyonic instability leads
  to the formation of regions with positive and negative  field values separated by domain wall configurations. These domains melt periodically at every field oscillation to reappear again just slightly distorted.
    \item  The time-dependent character of the symmetry-breaking minima prevents the usual stabilization around them, giving rise to a bi-modal  distribution pattern that survives for a large number of oscillations.
    \item Although the energy transfer to gradients is initially just a small fraction of the total energy density, it grows monotonically at the expense of the interaction and potential contributions. This leads to a final virialized state in which the kinetic energy density of the spectator field equals its gradient  counterpart.
    \item The virialization process is associated with a smooth turbulent regime, similar to that in Refs.~\cite{Khlebnikov:1996mc,Micha:2004bv,Micha:2002ey,Berges:2010ez}.
    The continuous population of  ultraviolet (UV) modes by this mechanism translates into an effective equation of state $w_\chi\simeq 1/3$. Combined with kination, this allows for the onset of radiation domination for \textit{general interacting  potentials}. This contrasts with the naive expectation following from the homogeneous approximation, where this transition is only possible for a quartic interaction \cite{Opferkuch:2019zbd}.
    \item Even if the late-time evolution of the spectator coincides with that of a radiation fluid, the associated spectrum of perturbations remains highly non-thermal for an extended period of time, precluding the naive extraction of thermodynamical quantities like temperature. 
\end{enumerate}
The paper is organized as follows. In Section \ref{sec:model}, we introduce the basic ingredients of the model, leaving for Section \ref{sec:linear} the description of the rapid amplification of the spectator field fluctuations through the Hubble-induced tachyonic instability. This classicalization process is essential for the analysis of the subsequent non-linear stage, which we study in Section \ref{sec:nonlinear} via $3+1$ classical lattice simulations. The implications of our findings for the onset of radiation domination are discussed in Section \ref{sec:implication}. Section \ref{sec:limitations} contains a summary of our main theoretical assumptions and future research lines, including a discussion on the possibility of identifying the spectator field with the Standard Model Higgs. Finally, our conclusions are presented in Section  \ref{sec:conclusions}.

\section{Non-minimally coupled spectator field}\label{sec:model}
A cosmological expansion epoch with a stiff equation-of state parameter ($w>1/3$) appears naturally in non-oscillatory quintessential inflation models aiming to provide a unified description of inflation and dark energy using a single degree of freedom \cite{Wetterich:1987fm,Wetterich:1994bg,Peebles:1998qn,Spokoiny:1993kt,Brax:2005uf,Wetterich:2013jsa,Wetterich:2014gaa, Hossain:2014xha,Agarwal:2017wxo,Geng:2017mic,Dimopoulos:2017zvq,Rubio:2017gty,Dimopoulos:2017tud,Akrami:2017cir,Garcia-Garcia:2018hlc}. In this type of scenarios, the absence of a potential minimum triggers generically a \textit{kination} or \textit{deflation} period immediately after the end of inflation. During this era, the total energy budget of the Universe is dominated by the kinetic energy density of the inflaton field, such that the global effective equation of state approaches $w\simeq 1$. Within this context, let us consider the following $Z_2$ symmetric action for an \textit{energetically subdominant spectator} field $\chi$,\footnote{In this \textit{spectator-field} approximation, there is no practical difference between analyzing the problem in a non-minimally coupled frame including \eqref{eq:lagchi} or in an Einstein frame in which the gravitational part of the full theory action takes the usual Einstein-Hilbert form. Indeed, by performing a Weyl transformation $g_{\mu\nu} \to \Omega^{-2} g_{\mu\nu}$ with $\Omega^2 =1 -\xi\chi^2/M^2_P$ one obtains, as expected, an Einstein-frame representation coinciding with the original one up to a set of highly suppressed higher-dimensional operators ${\cal O}( \chi/M_P)$ \cite{Markkanen:2017dlc}. Interestingly, this equivalence holds for all backgrounds and not only for homogeneous and isotropic ones. }
\begin{equation}\label{eq:lagchi}
S_\chi=\int d^4 x \sqrt{-g} \left[-\frac{1}{2}g^{\mu\nu}\partial_\mu\chi \partial_\mu \chi-\frac{1}{2}\xi R  \chi^2-\frac14 \lambda \chi^4 \right]\,,
\end{equation}
 where we have intentionally omitted a potential mass contribution which, if sufficiently small, will not play any essential role in the following discussions  \cite{Bettoni:2018utf,Bettoni:2018pbl,Bettoni:2019dcw}. Here $R$ stands for the Ricci scalar constructed out of the metric $g_{\mu\nu}$ and $\xi$ and $\lambda$ are two dimensionless coupling constants that we will assume to be positive-definite in what follows. 

The dynamics of the field $\chi$ is characterized by the Klein--Gordon equation,
\begin{equation}\label{eq:eq_chicov}
 \Box\, \chi -\xi R\chi-\lambda \chi^3=0\,,
\end{equation}
together with the energy density and pressure following from the $00$ and $ii$ components of the associated energy-momentum tensor,
\begin{equation}
T^{(\chi)}_{\mu\nu}  = 
  D_{\mu} \chi  D_{\nu} \chi - g_{\mu\nu} \left[
   \frac{1}{2}D^\rho \chi D_\rho\chi    
  + \frac{\lambda}{4} \, \chi^4 \right]  +\xi \Big[ G_{\mu\nu} 
  + \left(g_{\mu\nu} \Box -D_{\mu} \, D_{\nu}\right)
 \Big] \chi^2,    
  \label{eq:emt}        
\end{equation}
with $D_\mu$ the covariant derivative, $G_{\mu\nu}$ the Einstein tensor and $\Box=D^{\lambda} D_{\lambda}$ the d'Alambertian operator. Note that the contraction of $T^{(\chi)}_{\mu\nu}$ with arbitrary timelike vectors  $u^\mu$ and  $u^\nu$ is not guaranteed to be positive definite at all times \cite{Ford:1987de,Ford:2000xg,Bekenstein:1975ww,Flanagan:1996gw}, thus violating the weak energy condition $T^{(\chi)}_{\mu\nu}u^\mu u^\nu\geq 0$ within this particular sector. 

For a \textit{fixed} flat Friedmann--Lema\^itre--Robertson--Walker metric  $g_{\mu\nu}={\rm diag}(-1,a^2(t)\,\delta_{ij})$  with scale factor $a(t)$, we have
\begin{equation}\label{metricFLRW}
R=3(1-3w) H^2\,, \hspace{7mm} G_{00}=3H^2\,, \hspace{7mm} G_{ij}=3 \,w \,H^2 g_{ij}\,, \hspace{7mm}   \Box\chi =-\ddot \chi - 3H\dot \chi+\frac{\nabla^2}{a^2} \,,
\end{equation}
with $H=\dot a/a$ the Hubble rate, the dots denoting derivatives with respect to the coordinate time $t$, and $w$ the \textit{global} equation--of--state parameter. Using these expressions, we can rewrite Eqs.~\eqref{eq:eq_chicov} and \eqref{eq:emt} as 
\begin{eqnarray}
   &&  \hspace{-7mm} \ddot \chi + 3H\dot \chi-a^{-2}\nabla^2\chi +3\xi (1-3w) H^2 \chi+\lambda\, \chi^3=0\,, \label{eq:eq_chit} \\
  &&\hspace{-7mm}   \rho_\chi = T_{00}=\frac12 \dot\chi^2+\frac{1}{2a^2} \vert \nabla \chi\vert ^2 +\frac{\lambda}{4} \, \chi^4+ 3\xi\left[H^2+ H \partial_t -\frac{1}{3} 
\frac{\nabla^{2}}{a^{2}}\right]\chi^2 \,, \label{energychi} \\ 
  && \hspace{-7mm} p_\chi=\frac{1}{3a^2}\sum_i T_{ii} =\frac{1}{2}\dot{\chi}^{2}-\frac{1}{6a^2}\vert \nabla \chi\vert^2 - \frac{\lambda}{4} \, \chi^4+ 3\xi \left[   w H^{2}-\frac{1}{3} \left(\partial^2_t +2 H \partial_t  \right) +\frac{2}{9}\frac{\nabla^2}{a^2}\right]\chi^{2} \,.
  \label{pressurechi}  
\end{eqnarray}  
As  long  as  the  energy  density  in  the  scalar  field  is  small  and  the  corrections  to  the Planck mass are negligible, the background dynamics can be characterized independently of the spectator field. In particular, the equation-of-state parameter in the above expressions can be well-approximated by that of the quintessential inflation field. Having this in mind, we can distinguish two regimes:
\begin{enumerate}
    \item Inflation ($w=-1$): During this epoch, the Hubble-induced mass term in Eq.~\eqref{eq:eq_chit}  is positive definite, making the spectator field heavy for sufficiently large values of the non-minimal coupling ($\xi\gtrsim 1/12$). This confines the $\chi$ field to the origin of its effective potential and suppresses the generation of isocurvature perturbations \cite{Bettoni:2018utf,Bettoni:2018pbl}. 
    \item Kination ($w=1$): The onset of a kinetic dominated era soon after the end of inflation triggers the spontaneous symmetry breaking of the $Z_2$ symmetry and the subsequent evolution of $\chi$ towards large field values, $|\chi_{\rm min}|= \left(6\xi/\lambda\right)^{1/2}H$.~\footnote{Note that if the background equation of state were not stiff, this Hubble-induced mechanism would not operate at all.}
\end{enumerate}
To study the dynamics of the spectator field in between these cosmological epochs, we will assume the inflation-to-kination transition to be instantaneous as compared with the evolution of the scalar field and parametrize the temporal dependence of the scale factor and the Hubble rate as 
\begin{equation}\label{eq:backkin}
a= a_{\rm kin}
\left[1+3 H_{\rm kin} \left(t-t_{\rm kin}\right)\right]^{1/3}\,, \hspace{20mm}  H=\frac{H_{\rm kin}}{1+3H_{\rm kin}(t-t_{\rm kin})}\,, 
\end{equation}
with $a_{\rm kin}$ and $H_{\rm kin}$ the values of these quantities at the onset of kinetic domination, arbitrarily defined at $t=t_{\rm kin}$. Performing now the change of variables
\begin{equation}
\label{eq:newvar}
Y\equiv \frac{a}{a_{\rm kin}}\frac{\chi}{\,\chi_*}\,, \hspace{15mm}
\vec y \equiv \, a_{\rm kin} \chi_* \,\vec x \,,  \hspace{15mm}
z \equiv \, a_{\rm kin}\chi_* \, \tau\,,
\end{equation}
with $\chi_*\equiv \sqrt{6\xi} H_{\rm kin}$ and conformal time $\tau\equiv \int dt/a$, the spectator field action \eqref{eq:lagchi} becomes
\begin{equation}\label{actionchi}
S_\chi = \int d^3{\vec y} \,dz \,\left[\frac12(Y')^2 -\frac12\vert \nabla Y\vert ^2 \,+\, \frac12 M^2(z)Y^2-\frac{\lambda}{4}\, Y^4\right] \,,
\end{equation}
with the primes denoting derivatives with respect to the dimensionless conformal time $z$ and the spatial derivatives understood now as taken with respect to the new spatial coordinates $\vec y$. All the explicit temporal dependence in \eqref{actionchi} is included in the effective mass term
\begin{equation}
M^2(z)\equiv 
(1-6\xi)(\mathcal{H}^2+\mathcal{H}')=(4\nu^2-1)\mathcal{H}^2\,,
\end{equation}
constructed out of the comoving Hubble rate 
\begin{equation}\label{az}
\mathcal{H}(z)\equiv \frac{a'(z)}{a(z)}=\frac{1}{2(z+\nu)}\,, \hspace{15mm} \mathcal{H}'(z)=-2\mathcal{H}^2(z)\,,
\end{equation}
with
\begin{equation}
\nu\equiv \sqrt{\frac{3\xi}{2}}
\end{equation}
an integration constant ensuring that $z(t_{\rm kin})=0$. 
 In terms of the new variables, the time-dependent minima of the potential are located at
\begin{equation}\label{minimapos}
     \vert Y_{\rm min}\vert =  \left(\frac{4\nu^2-1}{\lambda}\right)^{1/2}\mathcal H(z)\,,
\end{equation}
 while Eqs.~\eqref{eq:eq_chit} and \eqref{energychi} become
\begin{equation}\label{EDOY}
Y''-\nabla^2\, Y-M^2(z) Y+ \lambda\, Y^3=0\,, \hspace{15mm} \rho_\chi = \mathcal{X}\,\rho_Y\,,\hspace{15mm} p_\chi = \mathcal{X}\,p_Y\,,
\end{equation}
with $\mathcal{X}\equiv\chi_*^4(a_{\rm kin}/a)^4$ an overall rescaling factor. Here 
\begin{equation}\label{eq:EEPPcomp}
\rho_Y\equiv K+G+V+ I_1 \,, \hspace{15mm} p_Y\equiv K-\frac{1}{3}G-V+I_2\,,
\end{equation}
stand respectively for the dimensionless energy density and pressure of the $Y$-field with
\begin{equation}\label{rhopY}
 K =\frac{1}{2}\left(Y'-\mathcal H Y\right)^2\,,\hspace{10mm}
G = \frac{1}{2}\vert\nabla  Y\vert ^2\,,\hspace{10mm}
V= \frac{\lambda}{4}Y^4, \\ 
\end{equation}
\begin{equation}
 I_1=2\nu^2\left[\mathcal H (Y^2)'- \mathcal H^2 Y^2-\frac{1}{3}\nabla^2 Y^2 \right]\,, \hspace{5mm }
I_2= 2\nu^2\left[\mathcal H (Y^2)'- \mathcal H^2 Y^2-\frac{1}{3} (Y^2)'' +\frac{2}{9}\nabla^2 Y^2 \right]\,, \label{I1} 
\end{equation}
the kinetic ($K$), gradient ($G$), potential ($V$) and interaction contributions ($I_1$,$I_2$) following from the corresponding terms in Eqs.~\eqref{energychi} and \eqref{pressurechi}. As anticipated after Eq.~\eqref{eq:emt}, the spectator  energy density is not positive definite for non-vanishing $\nu$. Note, however, that the \textit{total} energy density including the inflaton field and other potential matter components is positive definite at all times, as required by the $G_{00}$ component in \eqref{metricFLRW}.

\section{Linear evolution of the quantum system}\label{sec:linear}

Immediately after symmetry breaking, the action \eqref{actionchi} can be well approximated by that of a free scalar field with time dependent mass $M(z)$, making the problem exactly solvable. In this section, we follow the canonical formalism of Ref.~\cite{GarciaBellido:2002aj}, which has been applied to the present scenario in Ref.~\cite{Bettoni:2019dcw}, where we refer the reader for a more detailed discussion. Expanding the $Y$ field in Fourier modes with wavenumber
\begin{equation}
 \kappa\equiv  \vert \vec \kappa \vert\equiv \frac{\vert \vec k\vert }{a_{\rm kin}\chi_*} \,, 
\end{equation}
and introducing position and momentum operators satisfying the usual equal-time commutation relations $\left[Y_{\vec\kappa}(z),\Pi_{\vec \kappa'}(z)\right]=i \delta^3(\vec \kappa + \vec \kappa')$, the quantum Hamiltonian following from Eq.~\eqref{actionchi} in the Gaussian approximation becomes the sum of an infinite set of harmonic oscillators,
\begin{equation}
H = \frac{1}{2}  \int d^3\kappa\,\left[\Pi_{\vec \kappa}(z)\,\Pi^\dagger_{\vec \kappa}(z) +
 \omega_{\kappa}^2(z)\,Y_{\vec \kappa}(z)\,Y^\dagger_{\vec \kappa}(z)\right]\,,
\end{equation}
with time-dependent dispersion relation 
\begin{equation}\label{freq}
 \omega_{\kappa}^2(z)\equiv \kappa^2-M(z)^2\,.
\end{equation} 
As clearly illustrated by this expression, the tachyonic instability induced by kinetic domination affects \textit{not only} the zero mode of the scalar field but also all \textit{subhorizon} modes below the time-dependent mass $M^2(z)$, i.e., those within a sliding window $\kappa_{\rm min} (z)\lesssim \kappa \lesssim \kappa_{\rm max}(z)$, with 
\begin{equation} \kappa_{\rm min}(z)=\mathcal{H}(z) \quad \quad  {\rm and} \quad  \quad\kappa_{\rm max}=(4\nu^2-1)^{1/2}\kappa_{\rm min}\,.
\label{eq:kminkmax}
\end{equation}
The evolution of this Gaussian quantum system is determined by the Heisenberg equations,
\begin{equation}\label{eq:Heise}
\frac{d}{dz} \begin{pmatrix} \Pi_{\vec \kappa}(z) \vspace{2mm} \cr Y_{\vec\kappa}(z) \end{pmatrix}= 
\begin{pmatrix}
0 & -\omega^2_{\kappa}(z)\vspace{2mm}\cr 1 &0
\end{pmatrix} 
\begin{pmatrix}\Pi_{\vec\kappa}(z) \vspace{2mm} \cr Y_{\vec \kappa}(z)
\end{pmatrix} \,,
\end{equation}
being all physical information encoded in the two-mode correlation functions at equal or different times,
\begin{equation}\label{exp}
\langle  v^I_{\vec\kappa}(z)\, v^J_{\vec \kappa'}(z') \rangle = \Sigma^{IJ}_\kappa(z,z')\, \delta^3({\vec \kappa}+{\vec \kappa'}) \,,
\end{equation}
with $v_{\vec \kappa}\equiv \left(\Pi_{\vec \kappa}(z),Y_{\vec\kappa}(z)\right)^T$ and the brackets indicating a quantum average over the initial state. Using Eq.~\eqref{eq:Heise}, we can express the correlation matrix $\Sigma^{IJ}_\kappa(z,z')$ in \eqref{exp} in terms of its value at any reference time $z_r$, namely 
 \cite{Polarski:1995jg,Lesgourgues:1996jc,Kiefer:1998jk,GarciaBellido:2002aj} 
\begin{equation}\label{eq:correlationmatrix}
\Sigma_{\kappa}(z,z')={\mathbf M}_{\kappa}(z)\,\Sigma_{\kappa}(z_r,z_r)\, {\mathbf M}^T_{\kappa}(z')\,, \hspace{5mm} {\mathbf M}_{\kappa}(z)  =
\begin{pmatrix} \sqrt{\frac{2}{\kappa}}\, \textrm{Re}\, g_{\kappa}(z)&\sqrt{2\kappa}\, \textrm{Im}\, g_{\kappa}(z)\vspace{2mm}\cr  
-\sqrt{\frac{2}{\kappa}}\,\textrm{Im}\,f_{\kappa}(z) & \sqrt{2\kappa}\,\textrm{Re}\, f_{\kappa}(z)
\end{pmatrix} \,,
\end{equation}
with $f_\kappa$ a solution of the Schr\"odinger-like differential equation
\begin{equation}\label{eq:feom}
f_\kappa '' + \omega^2_\kappa(z)\,f_\kappa = 0\,, 
\end{equation}
with initial conditions $ f_\kappa(z_r)$ and $f'_{\kappa}(z_r)$ and  $g_\kappa\equiv i f'_{\kappa}$.  Restricting ourselves to the large $\nu$ limit  (cf. Section \ref{sec:limitations}) and assuming the system to be initially in a vacuum state with $ f_\kappa(z_r=0) = 1/\sqrt{2\kappa}$ and $ f'_{\kappa}(z_r=0)=-i\sqrt{\kappa/2}$, the solution of Eq.~\eqref{eq:feom} takes the compact form \cite{Bettoni:2019dcw}
\begin{equation}\label{eq:soltach}
    f_{\kappa}(z)=\sqrt{z+\nu}\left[A_{\kappa}\, \mathcal{J}_\nu(\kappa(z+ \nu))- B_{\kappa}\,  \mathcal{Y}_\nu(\kappa(z+\nu)) \right]\,,
\end{equation}
with  $\mathcal{J_\nu}$ and $\mathcal{Y_\nu}$ the Bessel's functions of the first and second kind,
\begin{equation}
A_{\kappa}= {\cal Y}_{\nu }(\kappa \nu) \,\delta \,, \hspace{15mm}
  B_{\kappa}= \,{\cal J}_{\nu }(\kappa
   \nu)\,\delta-\frac{f_{\kappa}(0)}{\sqrt{\nu} {\cal Y}_{\nu }(\kappa \nu)} \,, 
\end{equation}
and 
\begin{equation}
  \delta = \frac{\pi  f_{\kappa }(0)}{4\sqrt{\nu}}  \left[1-2 \nu  +2
  \nu\left(\kappa  \frac{ {\cal Y}_{\nu -1}(\kappa  \nu)}{{\cal Y}_{\nu }(\kappa \nu) }-\frac{f'_{ \kappa }(0)}{f_{\kappa }(0)}\right)\right]\,.
\end{equation}
Using these expressions, we can easily evaluate the equal-time correlation matrix 
\begin{equation}\label{Sigmaz}
\Sigma_{\kappa}(z,z)=
\begin{pmatrix} 
|g_{\kappa}(z)|^2 &  F_\kappa(z) - \frac{i}{2}\vspace{3mm}\cr
 F_\kappa(z) + \frac{i}{2} &  |f_k(z)|^2
\end{pmatrix}\,,
\end{equation}
with 
\begin{equation}\label{WKBphase}
F_\kappa(z)\equiv 
\frac12\,\langle  \Pi_{\vec \kappa}(z)\, Y_{\vec \kappa}^\dagger(z) +  Y_{\vec \kappa}(z)\, \Pi_{\vec \kappa}^\dagger(z)\rangle
=  {\rm Im}\, (f_\kappa^* g_\kappa)  
\end{equation}
a WKB phase directly related to the Heisenberg uncertainty principle
 \cite{Polarski:1995jg,Kiefer:1998jk,Lesgourgues:1996jc,GarciaBellido:2002aj}
\begin{equation}
\Delta Y_\kappa^2\,\Delta \Pi_\kappa^2 = \vert F_\kappa(z)\vert^2 + \frac14 \geq {\frac14}\, \Big\vert \langle [Y_\kappa(z),\ \Pi^\dagger_\kappa(z)]\rangle\Big\vert^2\,.
\end{equation}
For $\vert F_\kappa(z)\vert \gg 1$,
the commutator $\langle\{Y_\kappa(z),\ \Pi_\kappa^\dagger(z)\}\rangle$ in this expression exceeds significantly the anticommutator $\langle|[ Y_\kappa(z), \Pi_\kappa^\dagger(z)]|\rangle$,
\begin{equation}
\langle\{Y_\kappa(z),\ \Pi_\kappa^\dagger(z)\}\rangle \gg \langle|[ Y_\kappa(z), \Pi_\kappa^\dagger(z)]|\rangle =  \hbar\,,
\end{equation}
meaning that the properties of the system can be well-approximated by those of a classical random field (see Fig.~\ref{fig:F_xz_levels}). 

The equal-time correlation matrix \eqref{Sigmaz} allows the computation of several related quantities characterizing the Gaussian random field \cite{Bettoni:2019dcw}. Among those to be considered in the subsequent analysis, we can highlight the two-point correlation function obtained by Fourier transforming the diagonal entry $\vert f_\kappa(z)\vert^2$,
\begin{equation}\label{eq:size0}
\zeta(\vec y,z) \equiv \left\langle Y(\vec y,z)Y(0,z)\right \rangle = \int \frac{d^3\kappa }{(2\pi)^3} \, e^{i\vec \kappa \cdot \vec y}\, \vert f_\kappa(z)\vert^2\,.
\end{equation}
For a highly-symmetric spherical configuration and large $\nu$ values, the momentum integral in this expression can be computed explicitly, getting  \cite{Bettoni:2019dcw}
\begin{equation}\label{eq:size}
\zeta(r,z) \simeq \zeta(0,z)  \,G_1(\kappa_* r) \,,
\end{equation}
with 
\begin{equation}\label{Yrmsanal}
 \zeta(0,z)\equiv Y^2_{\rm rms}(z)= \left(\frac{2 \nu-1}{8\pi \nu}\right)^2    \left(1+\frac{z}{\nu}\right)^{2 \nu +1} \kappa_*^2(z)
\end{equation}
the square of the time-dependent dispersion determining the root mean-square perturbation (rms),   $\kappa_*(z)\equiv 2\sqrt{\nu+1}\, \kappa_{\rm min}(z)$ a typical momentum scale and
\begin{equation}\label{eq:shape}
G_1(\kappa_* r)\equiv \sqrt{\frac{\pi}{2}}\frac{1}{\kappa_* r} \exp\left(-\frac{1}{2}\kappa^2_* r^2\right) \text{erfi}\left(\frac{ \kappa_* r}{\sqrt{2}}\right)\,,
\end{equation}
a shape function, with $\text{erfi}$ the imaginary error function \cite{abramowitz+stegun}.

\section{Non-linear evolution of the classicalized system} \label{sec:nonlinear}

As we have seen in the previous section, the quantum evolution of the spectator field drives the low momenta modes in Fourier space into a highly occupied state where quantum averages can be well approximated by classical ones \cite{GarciaBellido:2002aj,Bettoni:2019dcw}. This allows us to study the subsequent non-linear evolution using just classical lattice simulations, instead of more involved $2$-particle irreducible techniques \cite{Cornwall:1974vz,Berges:2000ur,Arrizabalaga:2004iw}. 
\subsection{Lattice discretization}

Our approach relies on standard numerical methods common to lattice implementations of classical field dynamics and makes use of the LATfield2 library for parallel field theory simulations \cite{Daverio:2015ryl}. This framework has been widely used for the analysis  of topological defects and validated thoroughly \cite{Daverio:2015nva,Hindmarsh:2017qff,Lopez-Eiguren:2017dmc,Hindmarsh:2019csc,Hindmarsh:2021vih}. We discretize the classical nonlinear equations of motion in both space and time, solving them in a finite-volume cubic lattice of size $L$ and $N^3$ points uniformly distributed, with periodic boundary conditions $Y_{Njk} = Y_{0jk}$, $Y_{iNk}=Y_{i0k}$ and $Y_{ijN} = Y_{ij0}$. The lattice parameters $L$ and $N$ are chosen in such a way that all the relevant scales are sufficiently covered during the simulation time. In particular, the infrared (IR) lattice cutoff $\kappa_{\rm lattice, IR}=2\pi/L$ is chosen to be much smaller than the minimum value of the characteristic momentum scale during the tachyonic regime,  $\kappa_{\rm lattice, IR}\ll\kappa_*(z_{\rm nl})$,  while the UV lattice cutoff $\kappa_{\rm lattice, UV}= \sqrt{3}/2 N\, \kappa_{\rm lattice, IR}$ is taken to exceed the maximum momentum excited by the tachyonic instability, $\kappa_{\rm lattice, UV}\simeq {\rm few} \times \kappa_{\rm max}(0)$.

A set of classical initial realizations is then evolved using the discretized equations of motion, with the expansion of the Universe assumed to be kination dominated. This evolution is completely deterministic, being the random character restricted only to the initial conditions inherited from the quantum treatment in Section \ref{sec:linear}. The numerical integration is started at a fixed time $z_i$  sufficiently advanced to guarantee that a large fraction of the modes have become classical but well before the time $z_{\rm nl}$,
\begin{equation}\label{eq:zeff}
\left(1+\frac{z_{\rm nl}}{\nu} \right)^{\nu+1/2}\simeq \frac{8 \pi \nu^2}{(2\nu-1)\sqrt{3 \lambda (\nu+1)}}\left(1-\frac{1}{4\nu^2}\right)^{1/2}\,,
\end{equation}
 at which non linearities become relevant and backreaction starts to affect the analytical results \cite{Bettoni:2019dcw}.
\begin{figure}
    \centering
    \includegraphics[scale=0.7]{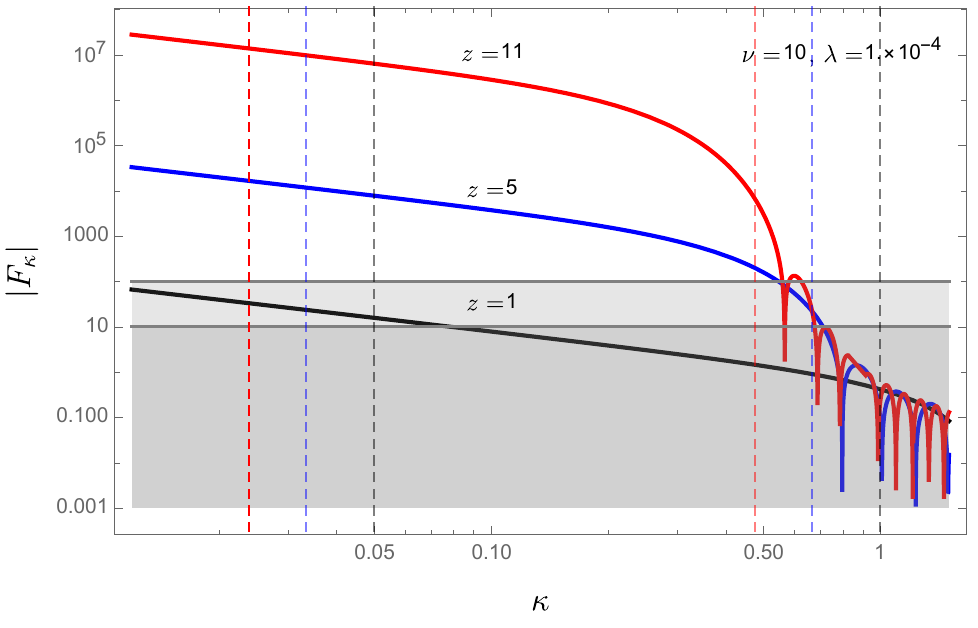}
    \caption{The amplitude of the WKB phase $\vert F_\kappa \vert$ in the correlation matrix \eqref{Sigmaz} as a function of the dimensionless momentum $\kappa$ evaluated at three different times. For the benchmark point  $\nu=10$, $\lambda=10^{-4}$, the quantum to classical transition takes place before the linear approximation ceases to be valid at $z_{\rm nl}(\nu,\lambda)= 11$. The shaded areas mark the regions $\vert F_\kappa \vert  \leq  10$ (darker) and $\vert F_\kappa \vert \leq 100$ (lighter). The dashed vertical lines indicate the values $\kappa_{\rm min}(z)$ and $\kappa_{\rm max}(z)$ in \eqref{eq:kminkmax}, with the same color coding as 
 $F_\kappa(z)$.}
    \label{fig:F_xz_levels}
\end{figure}
The amplitude of all modes that have not  become classical by the initialization time $z_i$ (namely those with $\vert F_\kappa \vert \lesssim 10$, cf. Fig.~\ref{fig:F_xz_levels}) is set to zero in our simulations. The remaining Fourier components are  initialized according to a Rayleigh distribution with uniform random phase $\theta_k \in [0,2\pi]$ and dispersion $ \sigma_k^2 = |f_k|^2$, i.e.
\begin{equation}
 P(|Y_k|)\,d|Y_k|\,d\theta_k =
\exp\Big(\!-{\frac{|Y_k|^2}{\sigma_k^2}}
\Big)\,{\frac{d|Y_k|^2}{\sigma_k^2}}\,{\frac{d\theta_k}{2\pi}}\,, \hspace{10mm} Y_k ' = \frac{F_k(z_i)}{ |f_k(z_i)|^2} \  Y_k\,.
\end{equation}
Since the short distance variations of the field fluctuations are not relevant for the thermodynamics description we will be interested in, we will make use of the ergodic postulate to identify ensemble averages of a given operator ${\cal O} (\vec y, z)$ with volume averages over the lattice volume $V=L^3$,
\begin{equation}
\langle {\cal O}(z)\rangle = \frac{1}{V}\int_{\,V} d^3 y\,\,  {\cal O} (\vec y, z)\,.
\end{equation}
In some cases, we will perform also suitable coarse-grainings on time with resolution $z_0$ \footnote{Note that $z_0$ is not necessarily a constant throughout the time evolution.} and/or averages over $J$ realizations of the stochastic initial conditions, 
\begin{equation}\label{coarsing}
\langle {\cal O}(z)\rangle =\frac1J \sum_{j=1}^{J}\frac{1}{z_0}\int^z_{z-z_0} dz\,\, \frac{1}{V}\int_{\,V} d^3 y\,\,  {\cal O}^{(i)} (\vec y, z)\,.
\end{equation}
Finally, we will define occupation numbers as \cite{Salle:2000hd,Salle:2000jb}
\begin{equation}\label{nkdef}
{n}_{\vec \kappa}(t) = \left(\vert \Pi_{\vec \kappa}\vert^2\,\vert Y_{\vec \kappa}\vert^2\right)^\frac12\,.
\end{equation}
For non-tachyonic frequencies this definition coincides with that following from the correspondence principle. Indeed, in the classical limit, the occupation number of a mode is given by  ${n}_{\vec \kappa}(t) =A/(2\pi)=\frac12 \vert \Pi_{\vec \kappa}\vert _{\rm max}|\vert Y_{\vec \kappa}\vert _{\rm max}$ with $A$ the area of the ellipse encircled by the trajectory in phase space. Taking into account that the modes are oscillating at late times and assuming them to be weakly coupled, we can write $|\Pi_{\vec \kappa}|_{\rm max} = (2\,\vert{\Pi}_{\vec \kappa}\vert^2)^{1/2}$ and $\vert Y_{\vec \kappa}\vert_{\rm max} = (2\vert Y_{\vec \kappa}\vert^2)^{1/2} $, recovering Eq.~\eqref{nkdef}.

\subsection{Results}

\begin{figure}[ht]
    \centering
    \includegraphics[scale=0.165]{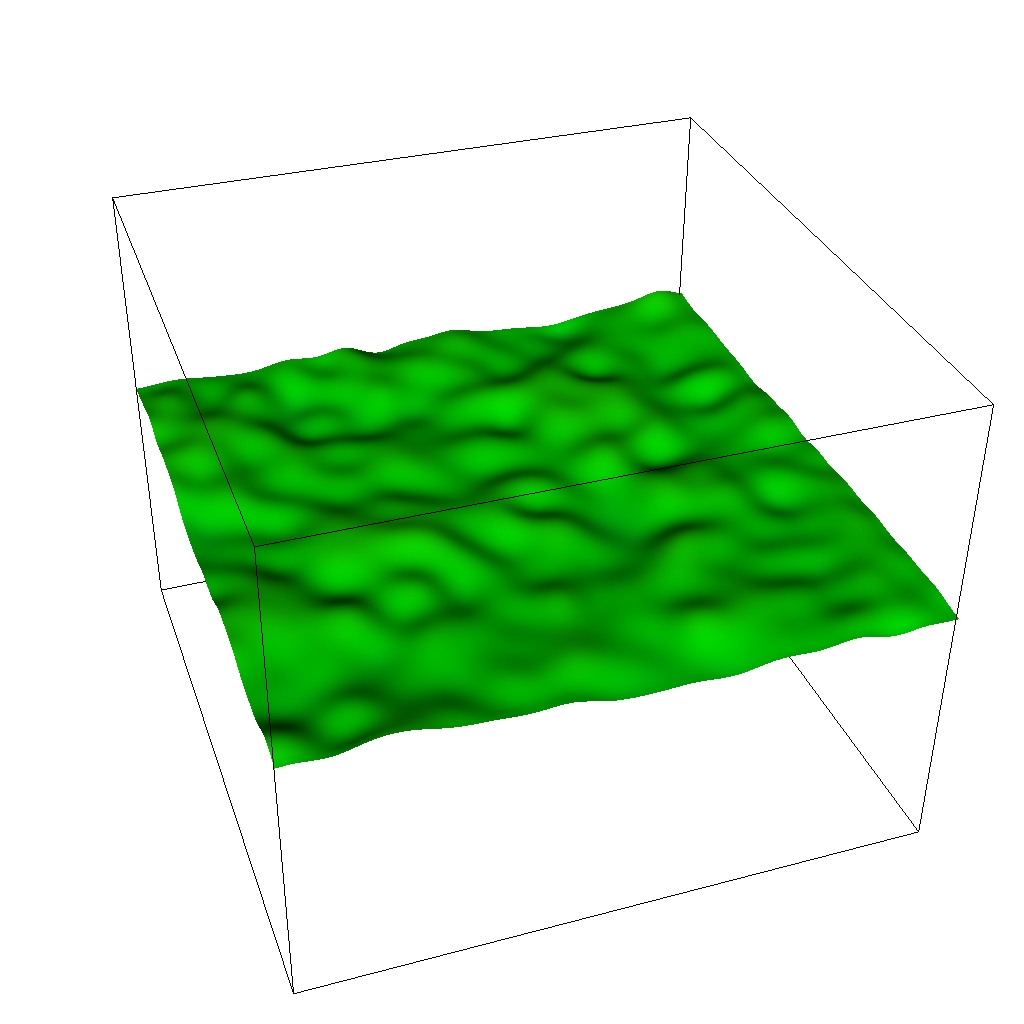}
    \includegraphics[scale=0.165]{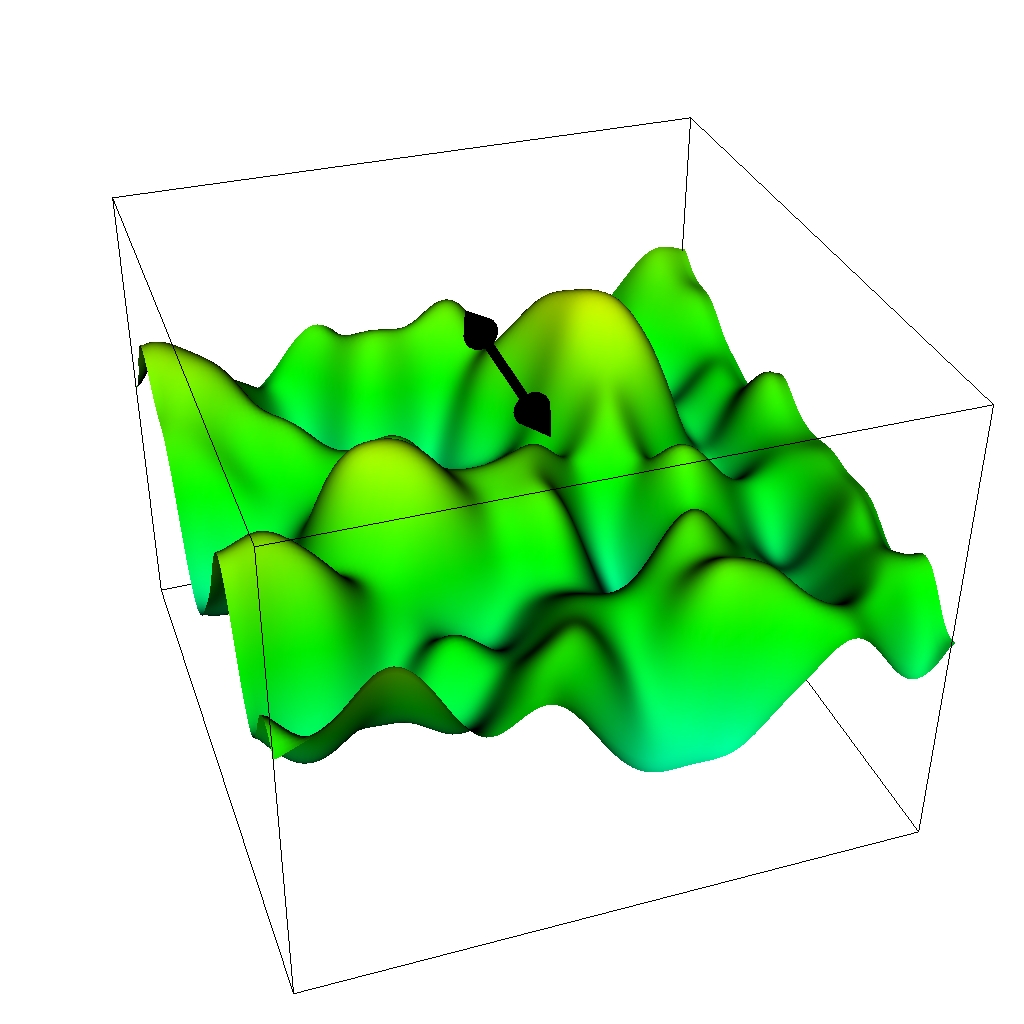}\\
    \includegraphics[scale=0.165]{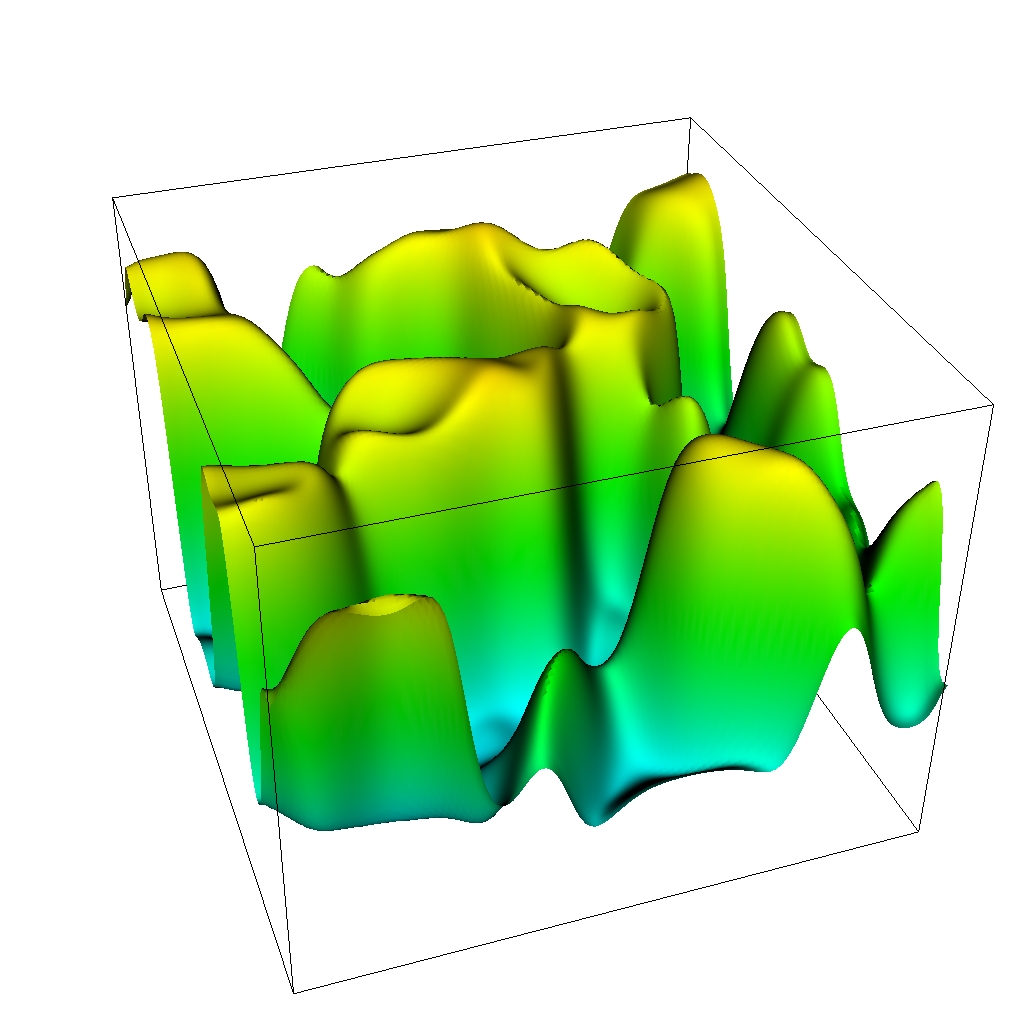}
    \includegraphics[scale=0.165]{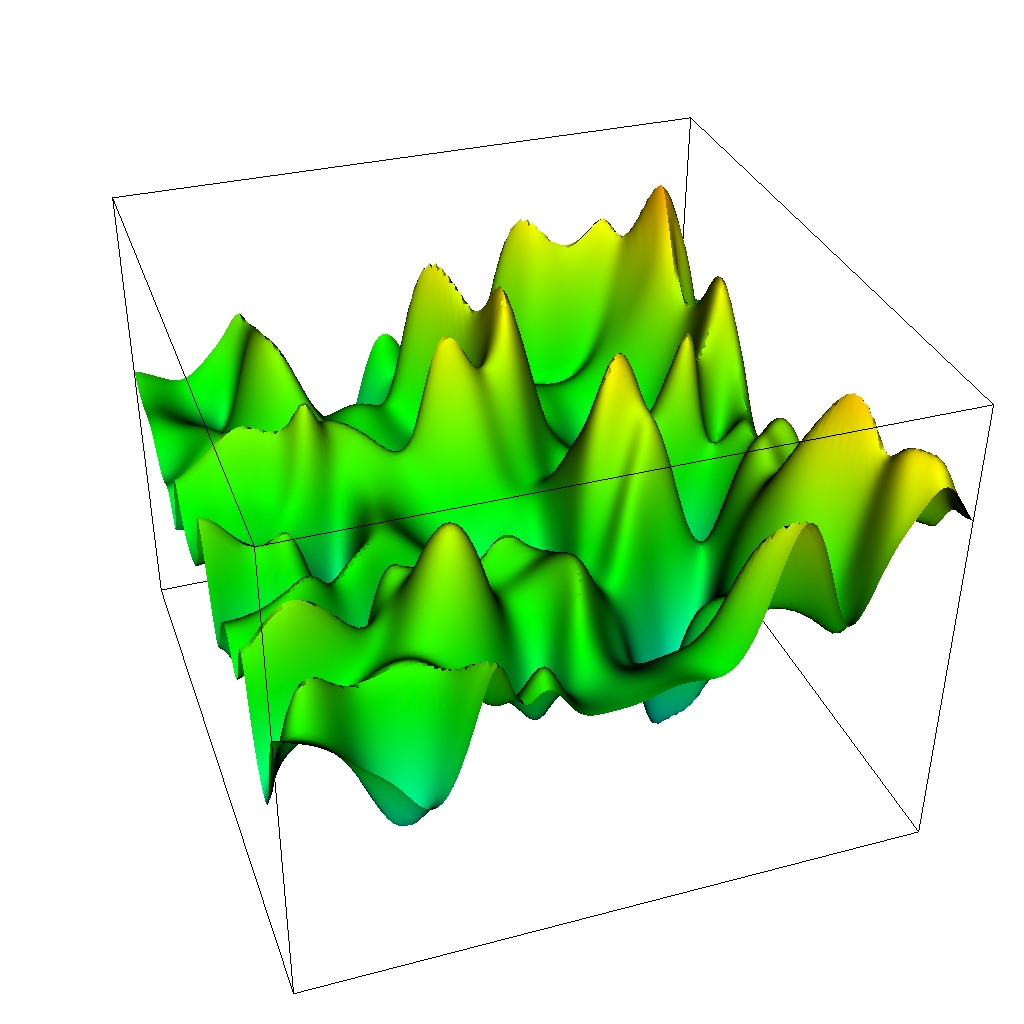}\\
    \includegraphics[scale=0.165]{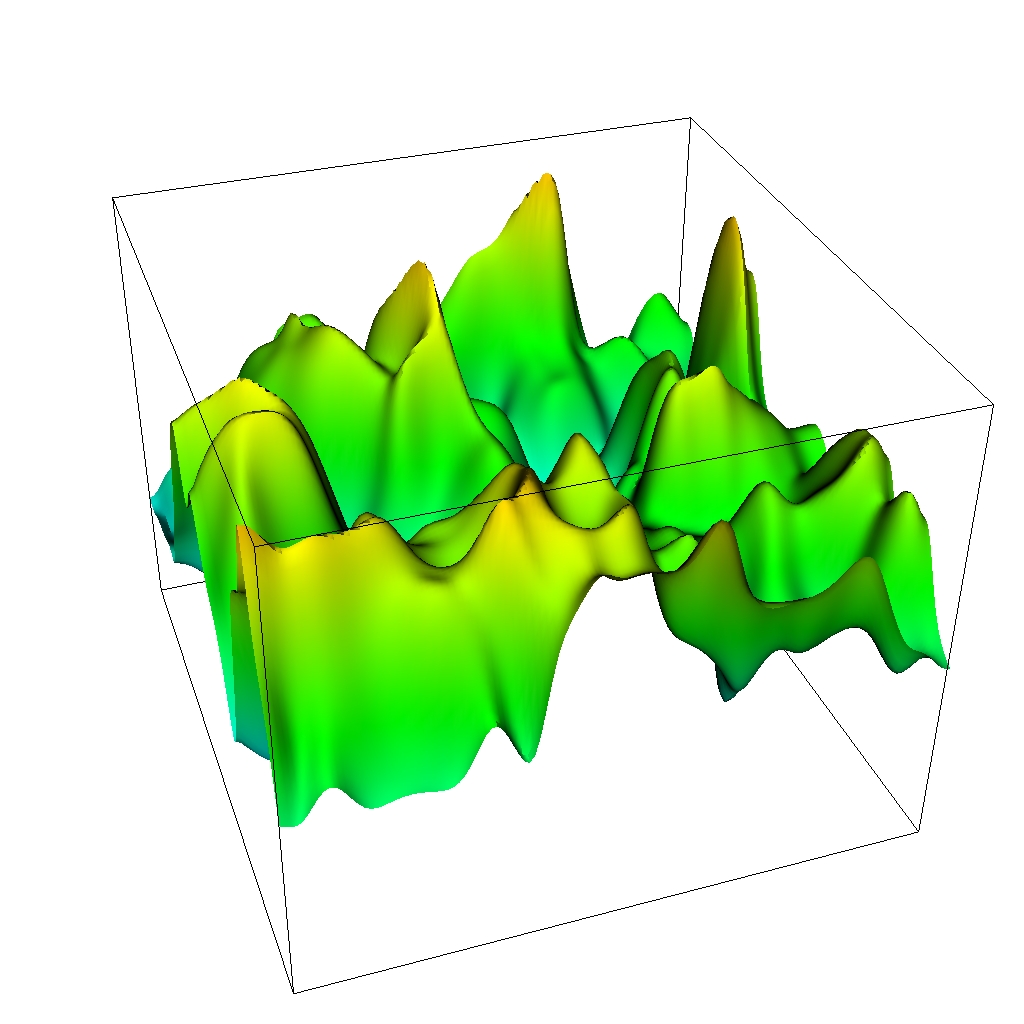}
    \includegraphics[scale=0.165]{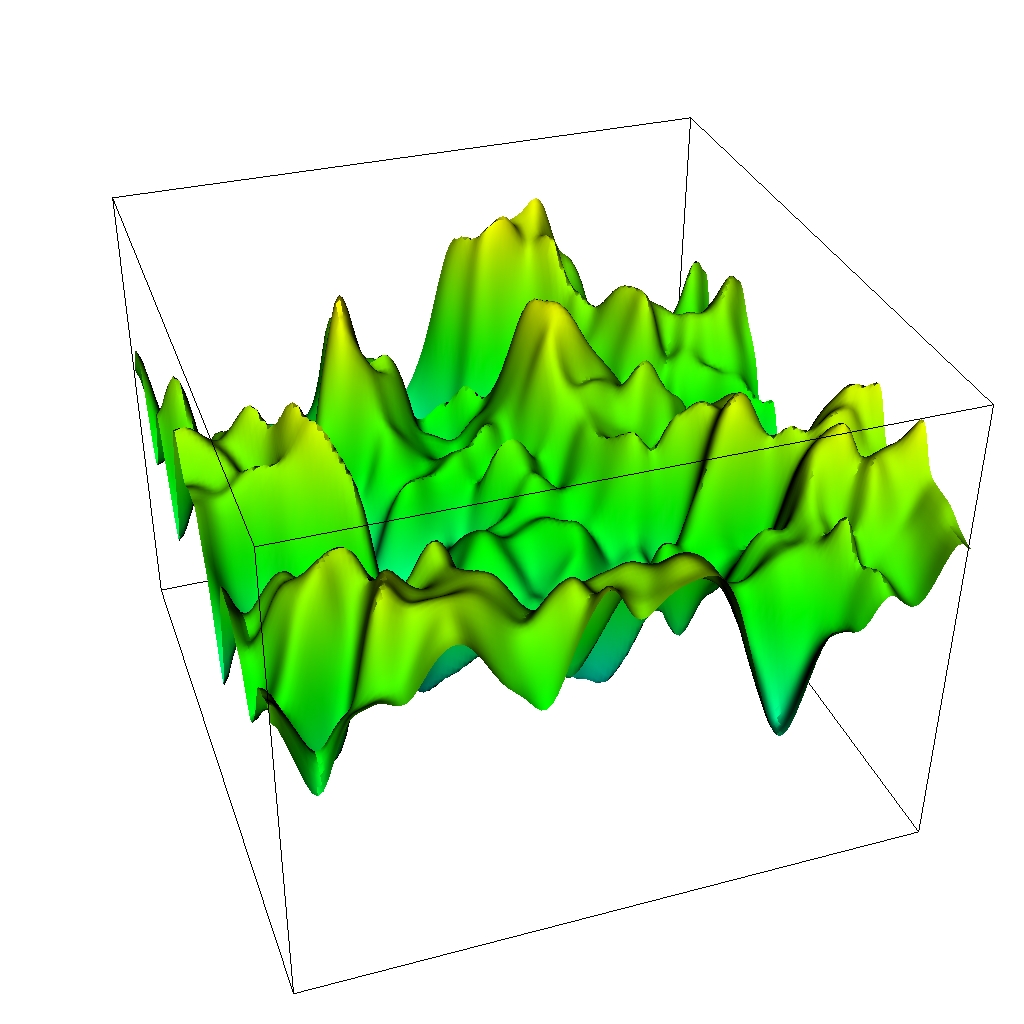}\\
    \includegraphics[scale=0.6]{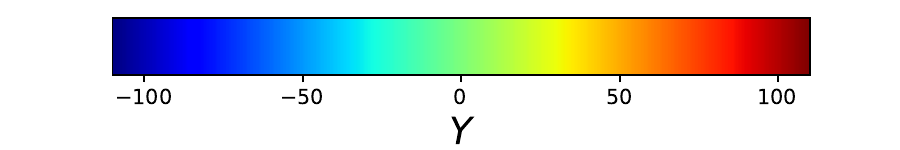}
    \caption{Spectator field growth for the benchmark point $\nu=10$, $\lambda=10^{-4}$. We display 2-dimensional sections of a fraction $V/8$ of the  3-dimensional simulation volume $V$ at times $z=5,\, 10,\, 15,\, 20,\,  25,  \, 60$, with the $z$-axis indicating the corresponding field value at each point in this surface. The black segment in the second panel indicates the typical scale of the highest peaks.}
    \label{fig:snapshots}
\end{figure}
\begin{figure}
    \centering
    \includegraphics[scale=0.185]{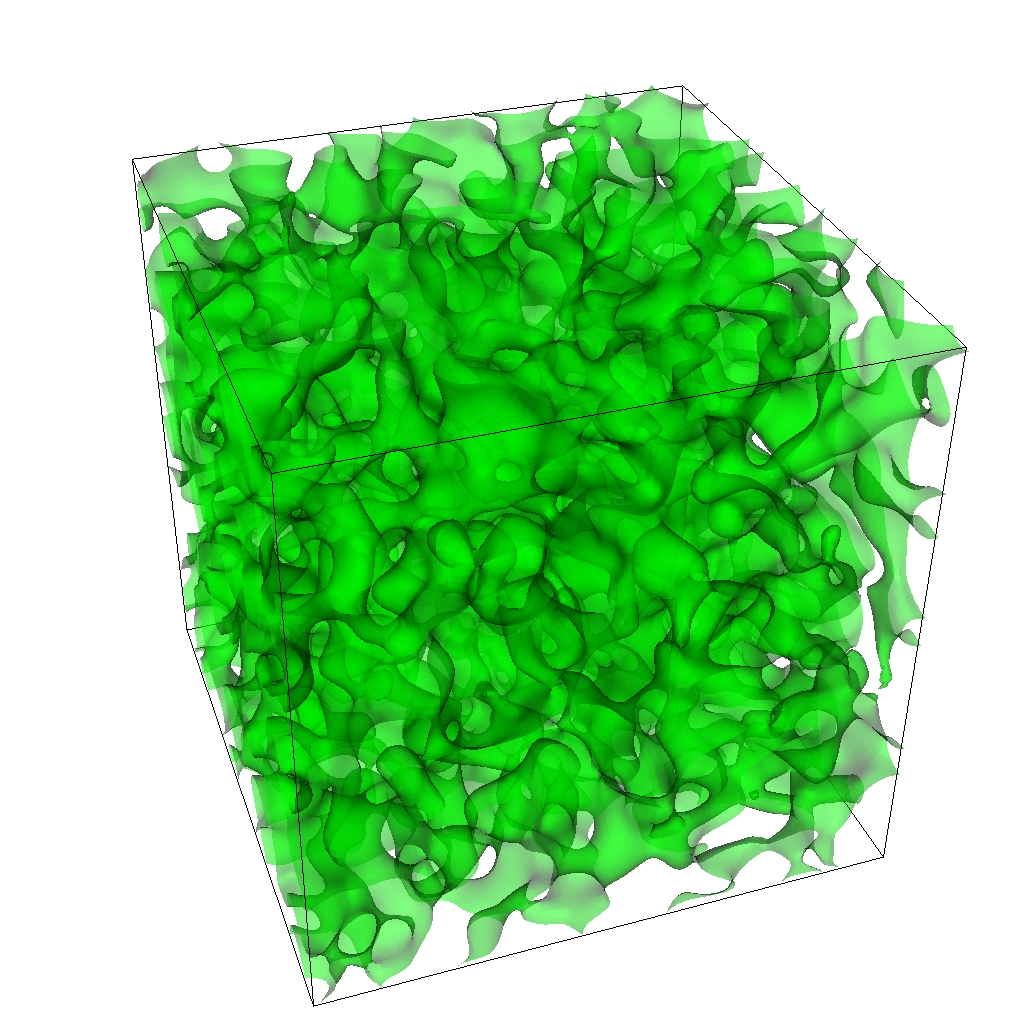}
    \includegraphics[scale=0.185]{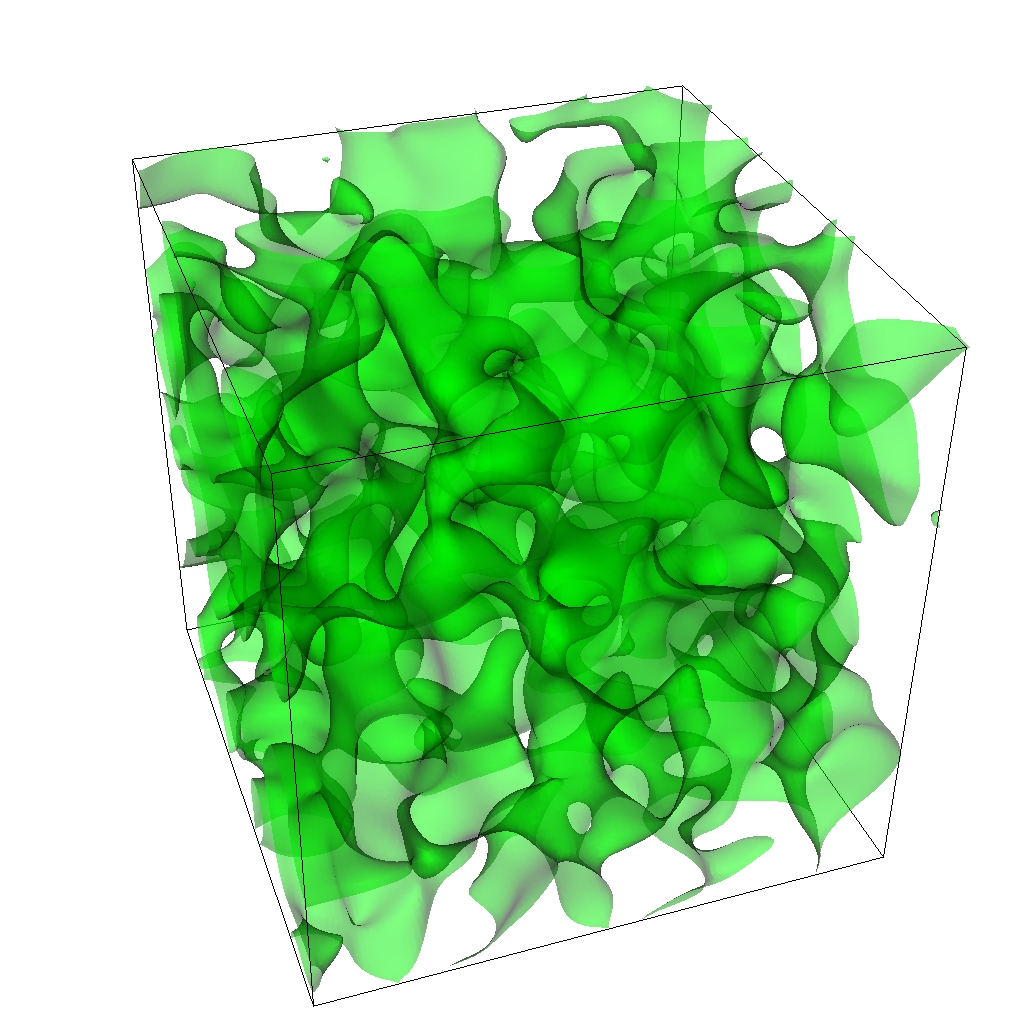}\\
    \includegraphics[scale=0.185]{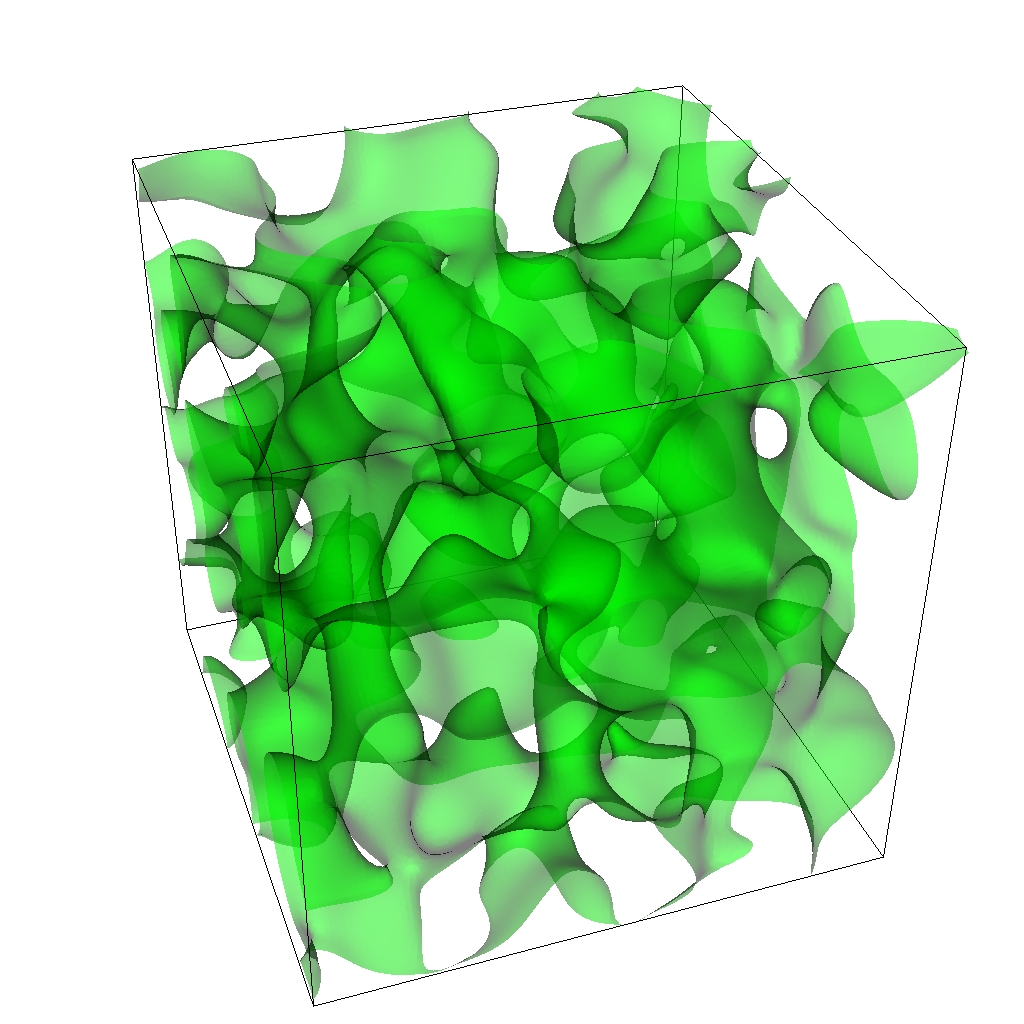}
    \includegraphics[scale=0.185]{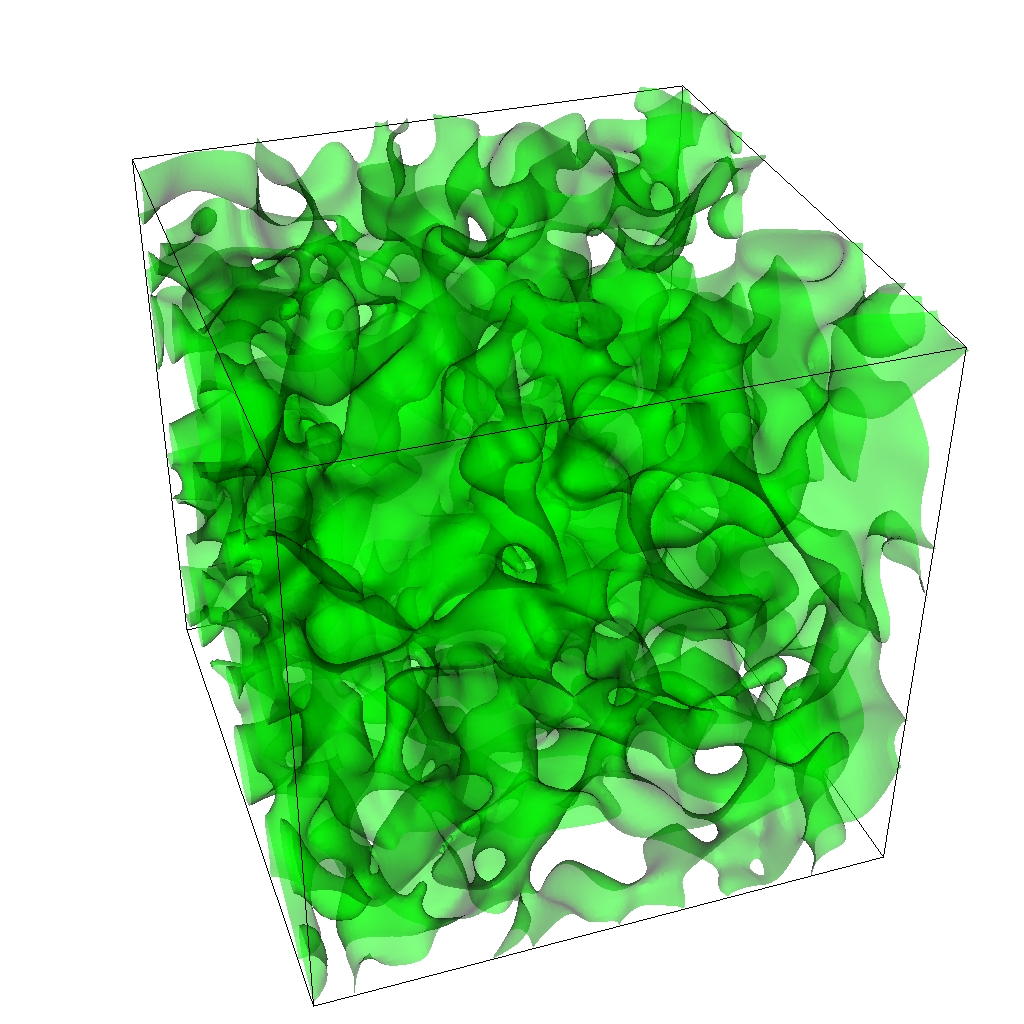}\\
    \includegraphics[scale=0.185]{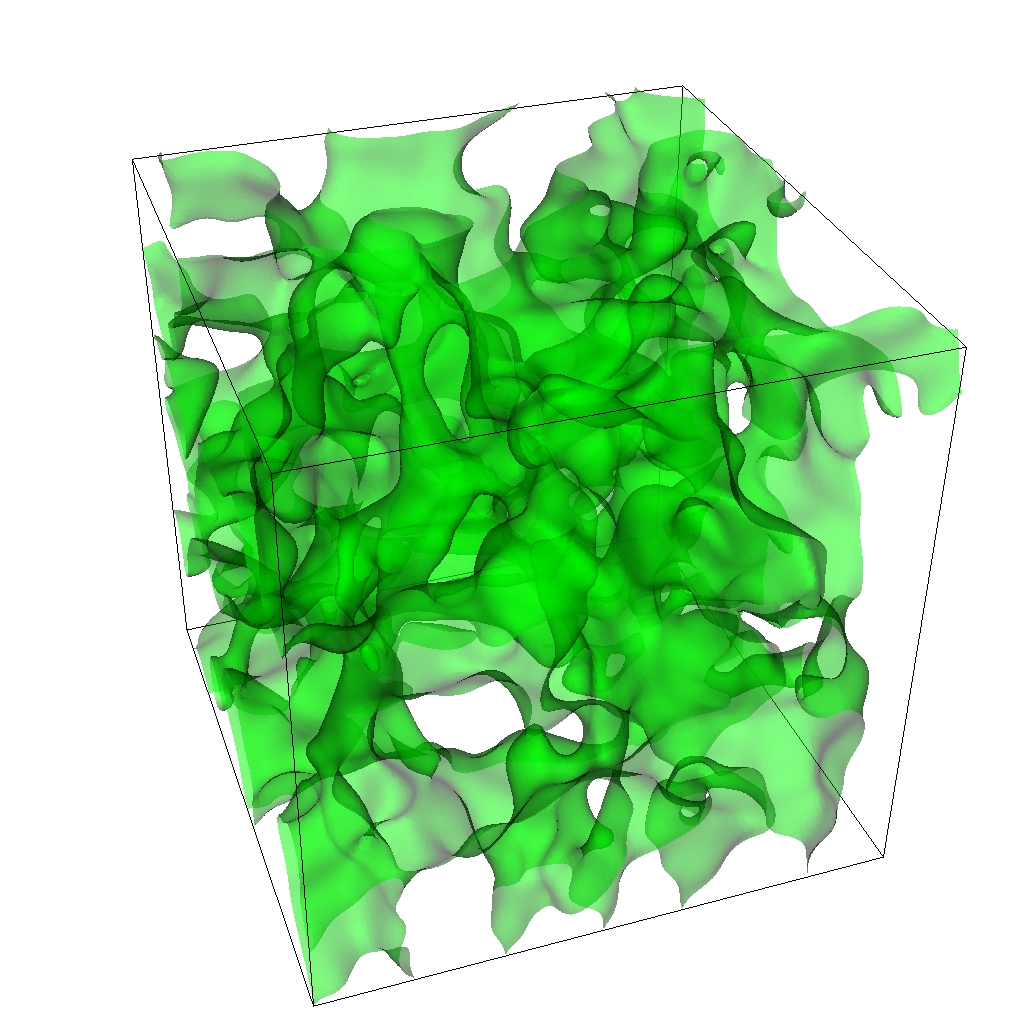}
    \includegraphics[scale=0.185]{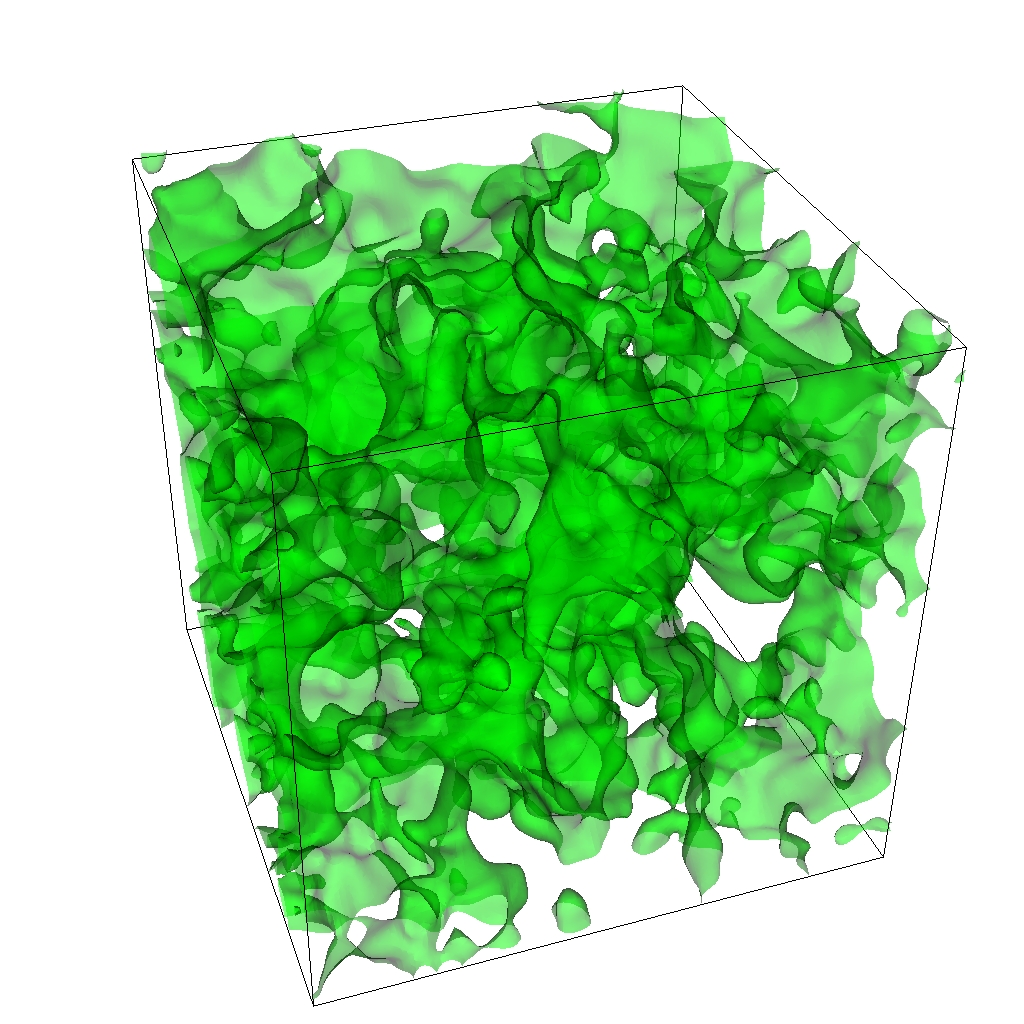}
    \caption{Domain wall configurations for the benchmark point $\nu=10$, $\lambda=10^{-4}$, obtained by plotting the 3-dimensional coordinates of the grid points with vanishing $Y$ at  times $z=5,\, 10,\, 15,\, 20,\,  25,  \, 60$. As before, we display only $1/8$ of the total simulation volume.}
    \label{fig:3DY0}
\end{figure}

To facilitate the comparison with the analytical results of Ref.~\cite{Bettoni:2019dcw}, and without lack of generality,  we concentrate in what follows on a benchmark point $\nu=10$, $\lambda=10^{-4}$, for which $z_{\rm nl}\simeq 11$. 

Our simulations are carried out on a lattice with $N=1024$ and $L=0.185 N$, which is initialized at $z=5$. With this choice of parameters the width of the domain walls is always larger than the lattice spacing, $dx=0.185$, making them resolvable throughout the entire simulation. The typical output is summarized in Fig.~\ref{fig:snapshots}, where we present several 2-dimensional slices of the 3-dimensional lattice volume at different times.\footnote{Computer generated movies covering the whole simulation volume can be found at  \href{https://youtube.com/playlist?list=PLl1K9-81ct6yHh6boyTiZAfNv_zlwux7-}{this URL}. } Although the initial conditions leading to these snapshots are random,  the qualitative features displayed there are universal up to small spatial and temporal shifts.
As anticipated above, an almost homogeneous Universe (1st panel) becomes rapidly separated into regions with positive and negative field values and typical size $L\lesssim \kappa_*^{-1}$, in agreement with the analytical result in Eq.~\eqref{Yrmsanal} (2nd panel and 3rd panel). These macroscopic regions are melted at every field oscillation (4th panel) to reappear again slightly distorted (5th panel). As time proceeds, additional fluctuations are generated at smaller and smaller scales, giving eventually rise to a rather diffuse structure  (6th panel). 
The domain walls separating regions of positive and negative $Y$ values can be explicitly detected. To this end, we plot in Fig.~\ref{fig:3DY0} the 3D coordinates of the grid points where $\vert Y\vert $ is close to zero. The displayed panels mirror the behaviour of Fig.~\ref{fig:snapshots}. In particular, we observe a resilient formation and dissolution of domains.  

A better characterization of the above symmetry-breaking picture can be achieved by considering the following observables:

\begin{enumerate}
    \item \textit{Probability distribution}: The total simulation volume containing the field value $Y$ at a given time $z$ is displayed in  Fig.~\ref{fig:histograms}. This probability distribution is initially peaked at $Y=0$, with a dispersion relation given by $\sigma_k^2 = |f_k|^2$. As time passes by, the distribution evolves towards the effective minima of the potential at large field values, eventually splitting in two. Contrary to other symmetry breaking scenarios \cite{Felder:2000hj,Felder:2001kt,Copeland:2002ku,GarciaBellido:2002aj}, the time-dependent character of the minima prevents the stabilization around them, giving rise to an oscillatory bi-modal pattern that survives for quite a number of oscillations. 
    \begin{figure}
    \centering
    \includegraphics[scale=0.4]{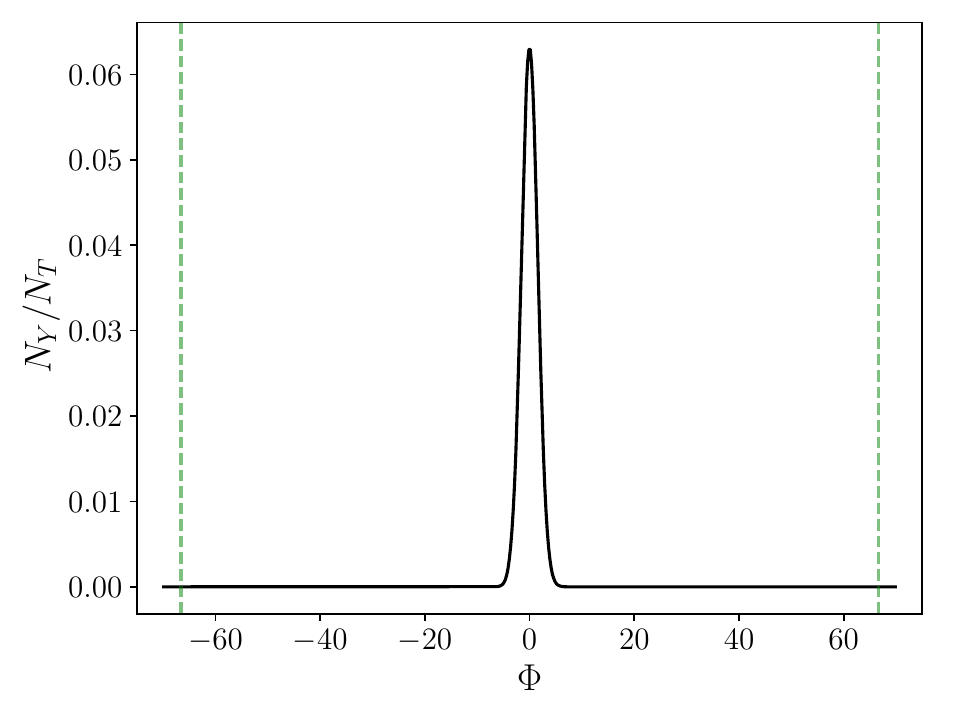}
    \includegraphics[scale=0.4]{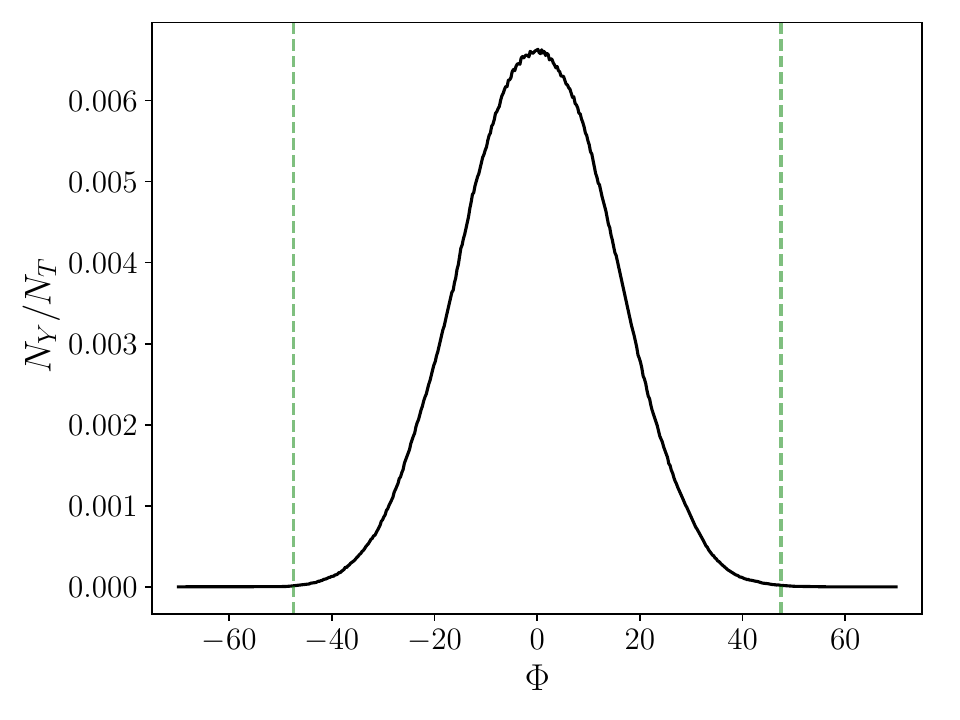}
    \includegraphics[scale=0.4]{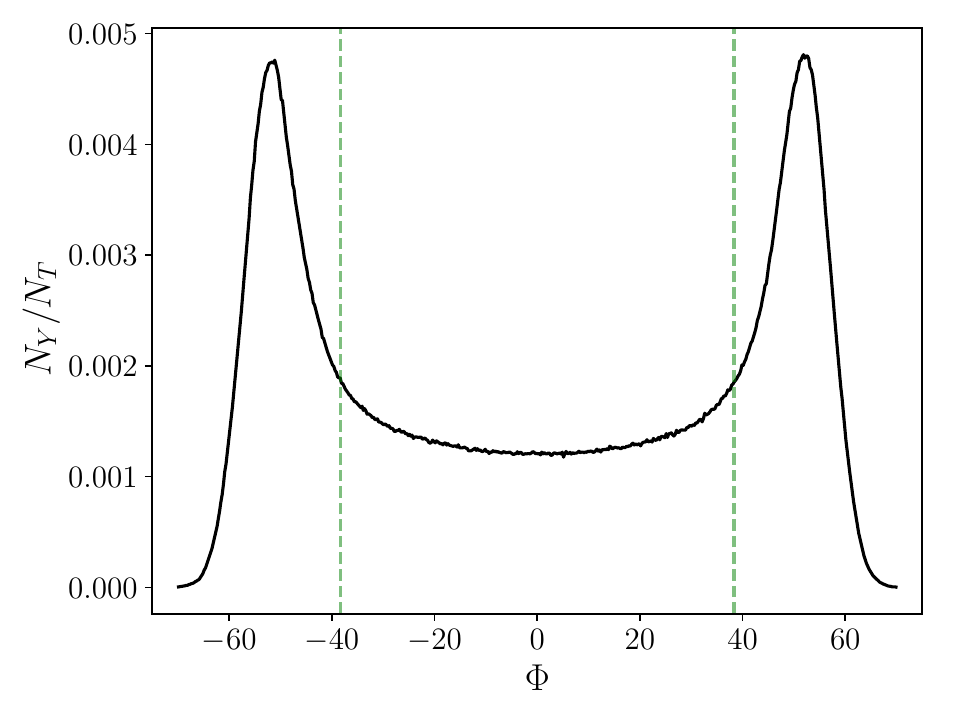}
    \includegraphics[scale=0.4]{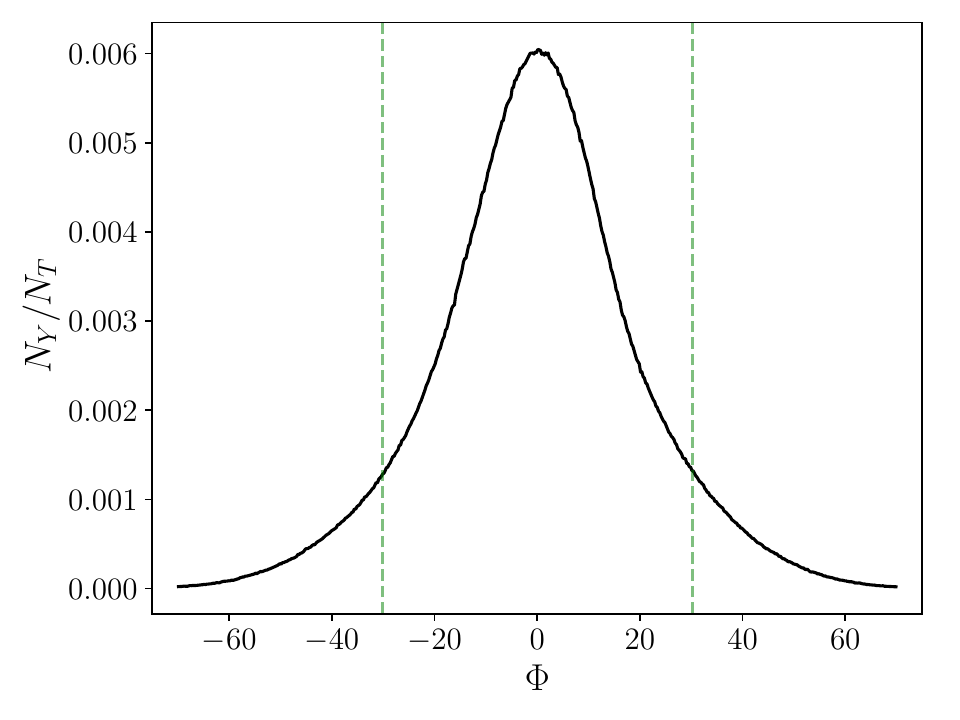}
    \includegraphics[scale=0.4]{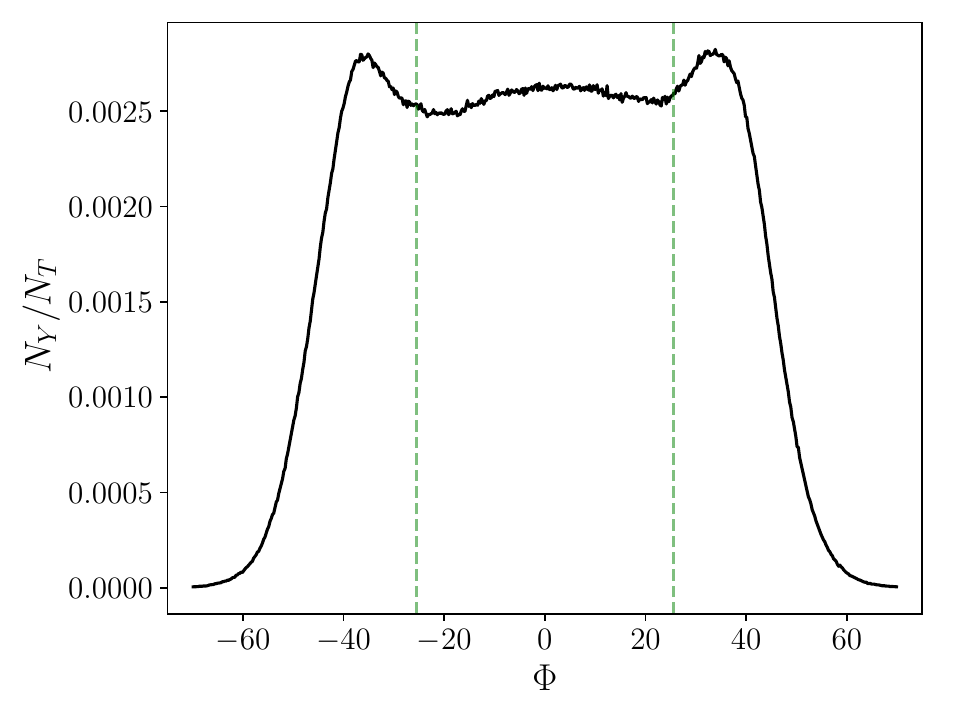}
    \includegraphics[scale=0.4]{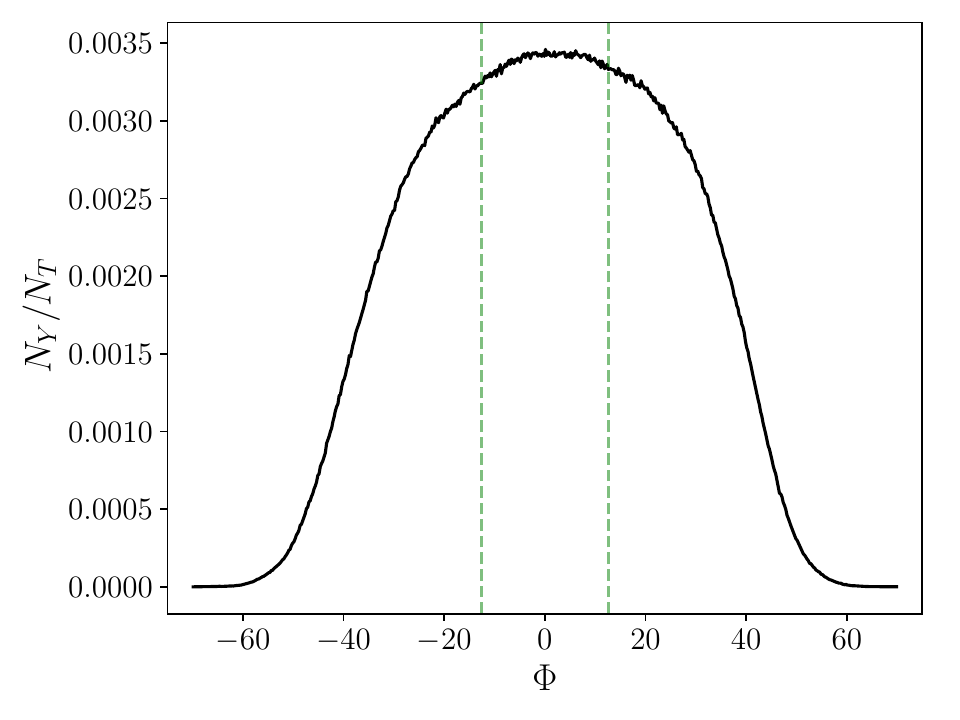}
    \caption{Evolution of the probability distribution after symmetry breaking for the benchmark point $\nu=10$, $\lambda=10^{-4}$. We display snapshots at times $z=5,\, 10,\, 15,\, 20,\, 25 ,\, 60$. 
    The vertical green-dashed lines signal the location of the time-dependent minima, as given in Eq~\eqref{minimapos}.}
    \label{fig:histograms}
\end{figure}
\item \textit{Time-dependent dispersion}: The root mean-square perturbation $Y_{\rm rms}$ is displayed in  Fig.~\ref{fig:meanfield} as a function of the dimensionless conformal time $z$ and the number of $e$-folds of kination,  $N_e\equiv \ln (a/a_{\rm kin})$. Note that, as argued in Refs.~\cite{Felder:2000hj,Felder:2001kt,Bettoni:2019dcw}, the picture of a homogeneously oscillating scalar field is incorrect when applied to spontaneous symmetry breaking. Indeed, when the initial value of the homogeneous component is close to zero, the classical rolling of the field is dominated by its quantum fluctuations, in agreement with the analytical result \eqref{Yrmsanal}.
    \begin{figure}
    \centering
    \includegraphics[scale=0.6]{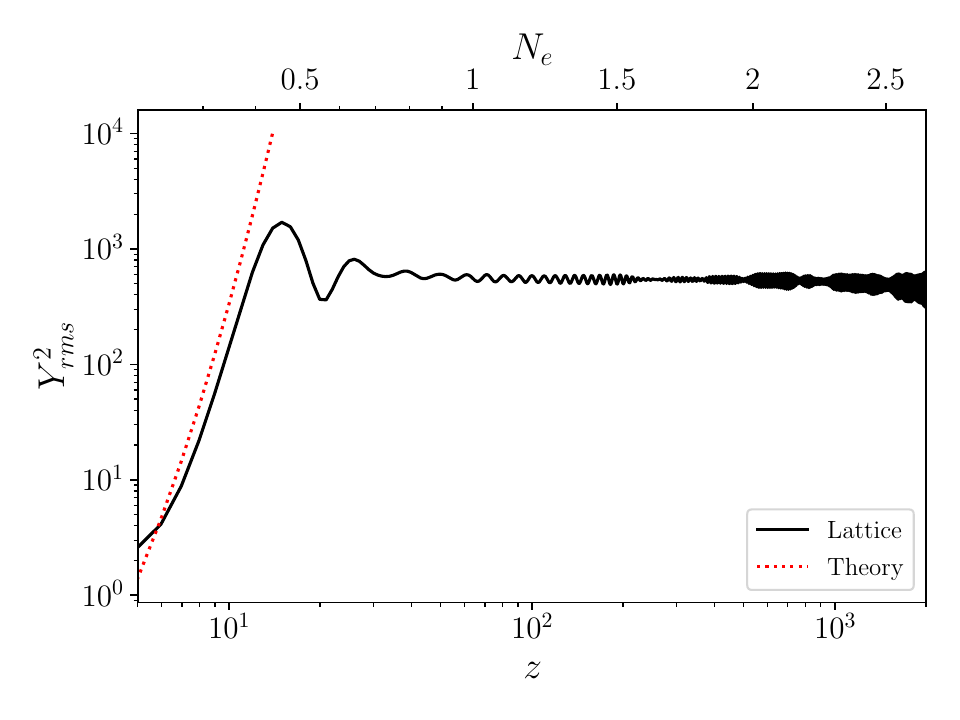}
    \caption{Evolution of the volume-averaged lattice $Y_{\rm rms}$ for the benchmark point $\nu=10$, $\lambda=10^{-4}$ as compared to the analytical result \eqref{Yrmsanal}, with $N_e$ the number of $e$-folds of kination.}
    \label{fig:meanfield}
\end{figure}
    \item \textit{Spectral distributions}:
The evolution of the occupation numbers \eqref{nkdef} is shown in Fig.~\ref{fig:occnum}. We observe a rapid growth of infrared fluctuations at the early stages of symmetry breaking followed by a slower motion towards the UV due to the potential interactions. Note that the displayed behaviour is far from thermal. Indeed, thermal equilibrium can be only achieved once the occupation numbers in the infrared have significantly decreased to $n_\kappa\sim \lambda^{-1/4}$; a limit that notably exceeds our simulation time. On top of that, it  is  well  known  that classical lattice simulations  cannot completely describe the approach  to equilibrium  due  to  the  Rayleigh-Jeans divergence\footnote{In the continuum temperature goes to zero; on the lattice it is cut-off dependent.} \cite{Berges:2013lsa}.

A simple characterization of the UV cascade can be obtained by computing the average wavenumber \cite{Destri:2004ck}
\begin{equation}\label{avermomentum}
\kappa_{c}\equiv \frac{\int_{\kappa>\kappa_0} d^3\kappa \,\vert \kappa\vert \, \vert \Pi_{\vec \kappa}\vert^2}{ \int_{\kappa>\kappa_0}  d^3\kappa \, \vert \Pi_{\vec \kappa}\vert^2} \,,
\end{equation}
where we have intentionally excluded the infrared modes filled at the onset of the simulation, namely those below $\kappa_0=0.8$.
As shown in Fig.~\ref{fig:momentum}, the momentum scale $\kappa_c$ is always smaller than the UV lattice cutoff, confirming that all relevant scales are well resolved throughout the entire simulation time.

\begin{figure}
    \centering
    \includegraphics[scale=0.55]{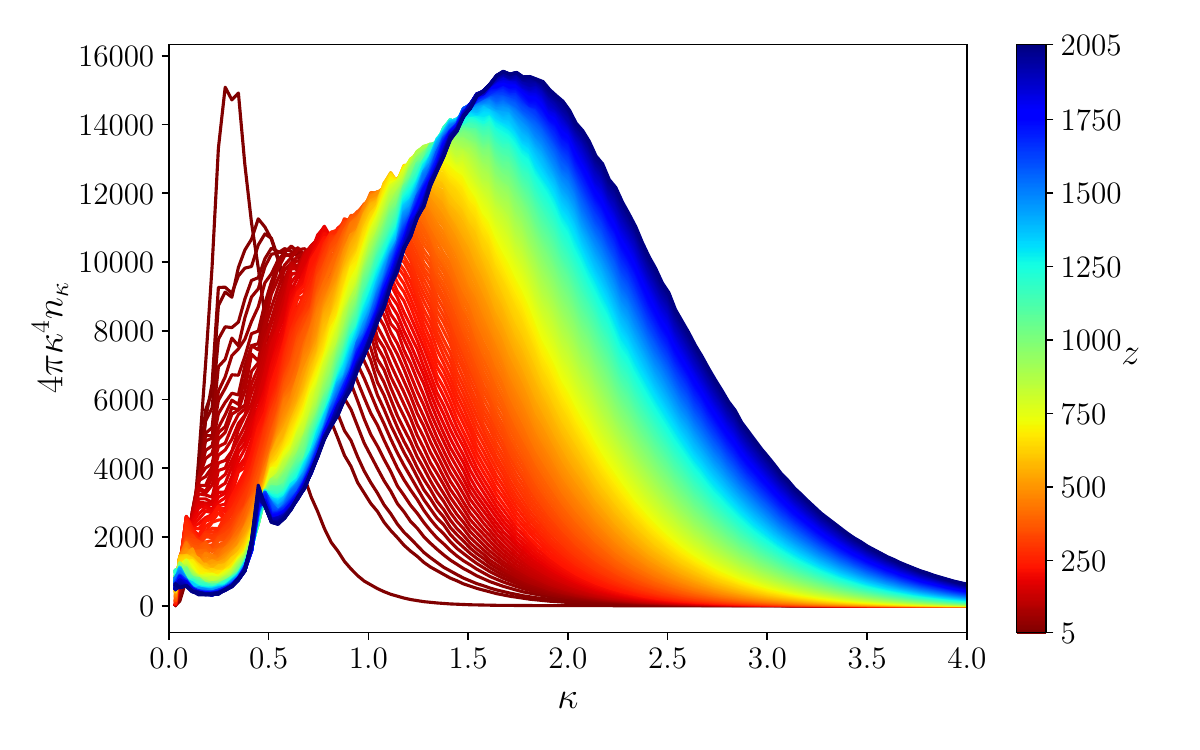}
    \caption{Time evolution of the occupation numbers \eqref{nkdef} for the benchmark point  $\nu=10$, $\lambda=10^{-4}$. Different colors correspond to different times, as indicated in the figure.}
    \label{fig:occnum}
\end{figure}
\begin{figure}
    \centering
    \includegraphics[scale=0.6]{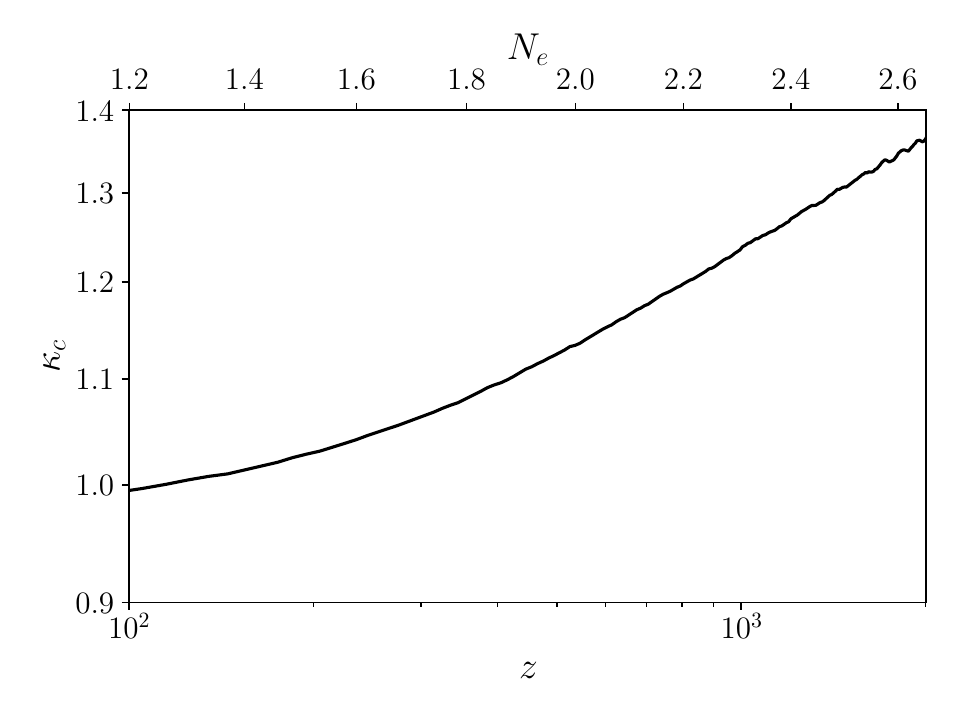}
    \caption{Time evolution of the UV average momentum \eqref{avermomentum}, with $N_e$ the number of $e$-folds of kination. }
    \label{fig:momentum}
\end{figure}

   \item \textit{Energy budget}: The evolution of the  kinetic, gradient, potential and interaction contributions to the total spectator energy density is displayed in Fig.~\ref{fig:energies}. After a rapid growth at early times and a violent oscillatory regime completely dominated by the interaction term, the system approaches an asymptotic state characterized by a large kinetic energy density. The smooth UV cascade in Fig.~\ref{fig:occnum} translates then into a steady flow of energy from the interaction and potential terms to a slowly growing gradient component. As time passes by, the system relaxes towards a virialized state with  $\braket{\textrm{K}}\simeq \braket{\textrm{G}}+ 2\braket{\textrm{V}}$, in agreement with the expectation for a quartic potential \cite{Boyanovsky:2003tc,Destri:2004ck}. 
\begin{figure}
    \centering
 \includegraphics[scale=0.6]{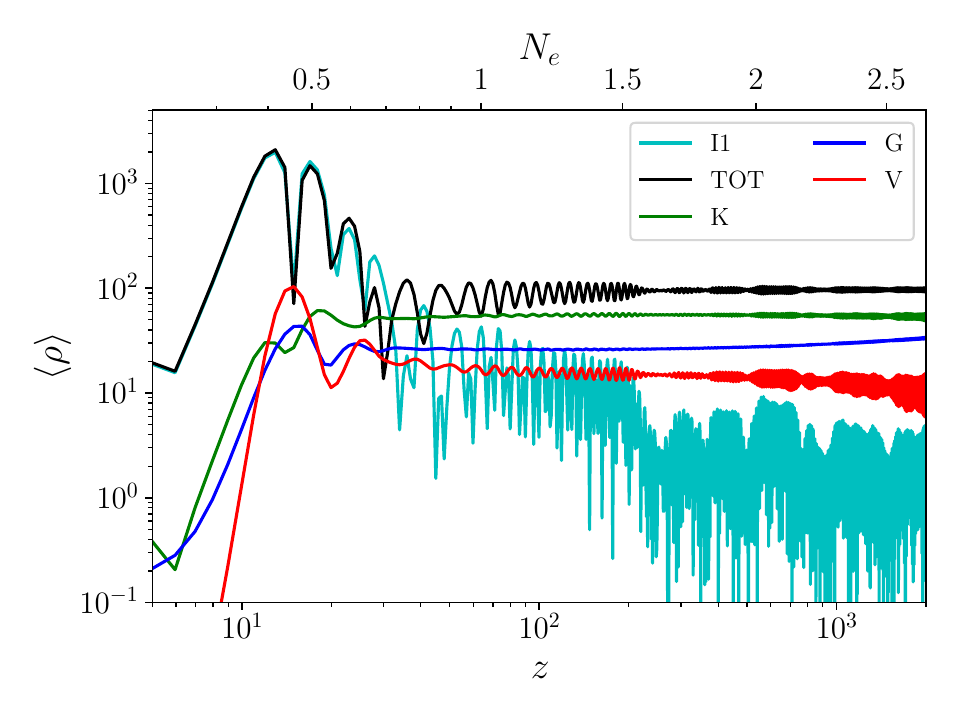}
    \caption{Evolution of the volume-averaged kinetic, gradient, potential and interaction contributions to the total spectator energy density for the benchmark point $\nu=10$, $\lambda=10^{-4}$, with $N_e$ the number of $e$-folds of kination. In order to facilitate the comparison, we display the \textit{absolute value} of the interaction term $I_1$, since this is not positive definite at all times.
    }
    \label{fig:energies}
\end{figure}  
    \item \textit{Equation-of-state parameter}: 
    The behaviour of the different energy density contributions is reflected also on the evolution of the associated equation-of-state parameter \begin{equation}\label{eq:wchi}
w_\chi\equiv \frac{\braket{p_\chi}}{\braket{\rho_\chi}}\,,
    \end{equation}
    which we display in Fig.~\ref{fig:eq_state}.  
    Since this quantity is rapidly oscillating as compared to the characteristic expansion rate, we perform here an additional coarse-graining in time, along the lines of Eq.~\eqref{coarsing}. As shown in the left panel of the figure, the averaged equation of state is initially negative, becoming positive only during the tachyonic-to-oscillatory transition. 
  This unusual behaviour is completely determined by the interactions terms in \eqref{I1} (see also Fig.~\ref{fig:energies}) and completely absent in standard minimally-coupled scenarios, where the equation of state is always positive definite.  The small discrepancy between the lattice and the analytical estimate in Appendix \ref{sec:appendix1} can be traced back to the fact that the latter assumes exact homogeneity and complete irrelevance of the non-linear potential terms while, albeit small initially, gradients are always present in the simulation and the potential terms can contribute. As shown in the right panel, $w_\chi$ approaches  a value rather close to that of a radiation fluid ($w_\chi=1/3$) at late times. This behaviour is intrinsically related to the virialization process displayed in Fig.~\ref{fig:energies} and completely independent of whether the system is already thermalized or not \cite{Podolsky:2005bw}.

\begin{figure}
    \centering
    \includegraphics[scale=0.45]{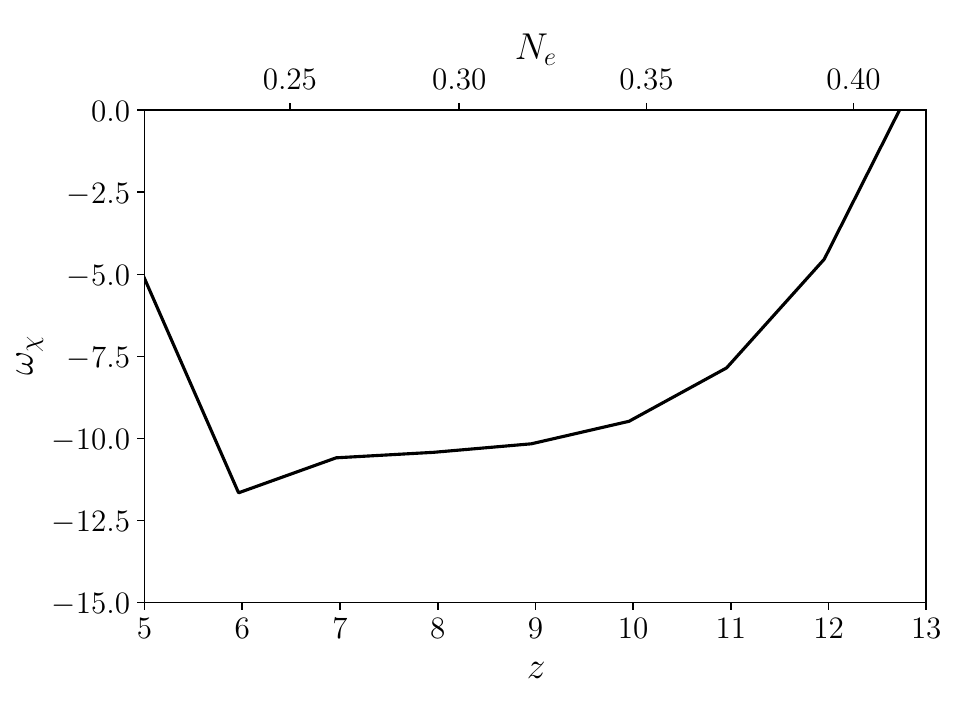}
    \includegraphics[scale=0.45]{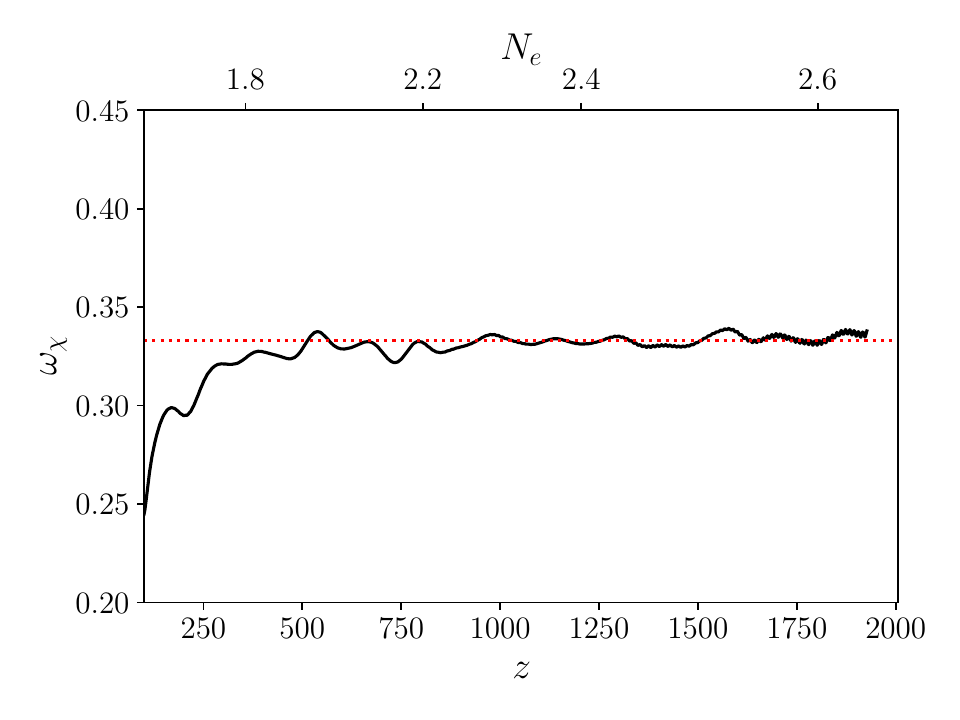}
   \caption{(Left) Evolution of the volume- and time-averaged equation-of-state parameter \eqref{eq:wchi} immediately after symmetry breaking, with $N_e$ the number of $e$-folds of kination. 
   (Right) Evolution of the same quantity at the latest stages of our simulation. The red horizontal line corresponds to the radiation value $1/3$.}
    \label{fig:eq_state}
\end{figure}
    \item \textit{Non-Gaussianities}: To characterize the level of non-gaussianity in our scenario, we consider the so-called \textit{excess kurtosis}, customarily defined as
    \begin{equation}\label{excessK}
\Delta K\equiv \frac{\langle Y^4\rangle}{\langle Y^2\rangle^2}-3\,.
    \end{equation}
 
\begin{figure}
    \centering
    \includegraphics[scale=0.6]{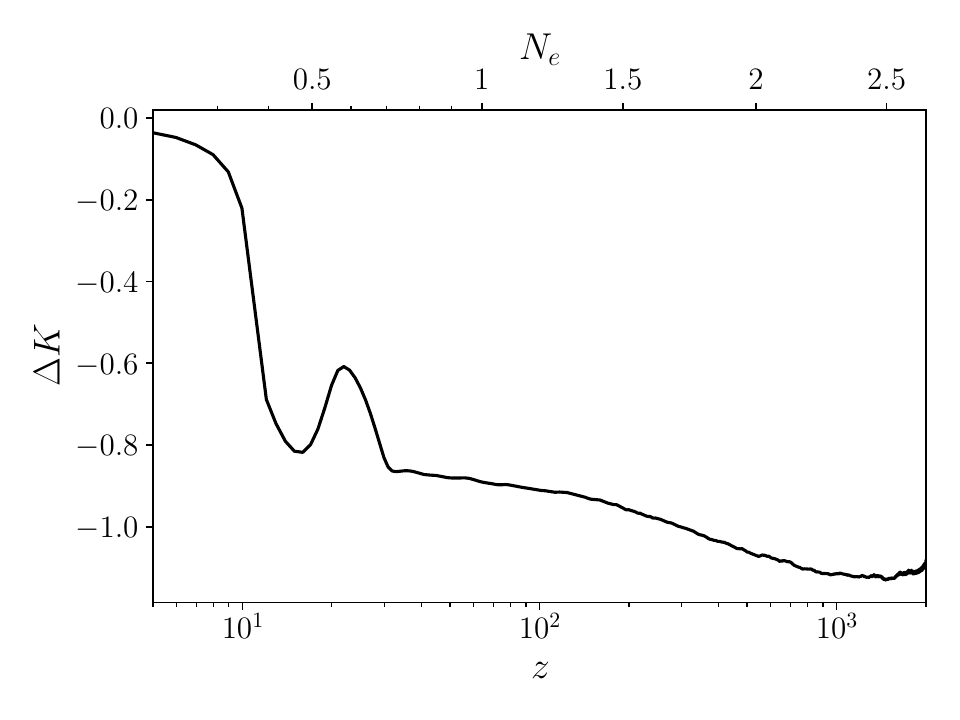}
    \caption{Time-averaged evolution of the \textit{excess kurtosis} \eqref{excessK} for the benchmark point $\nu=10$, $\lambda=10^{-4}$, with $N_e$ the number of $e$-folds of kination. }
    \label{fig:noGauss}
\end{figure}
\end{enumerate}
The above results should be understood as robust against variations of the initializing time $z_i$ and the lattice cutoffs. In particular, the temporal evolution of the different energy components is almost insensitive to the chosen initial condition and dominated by the modes lying far from the edges, where cutoff effects are subdominant.  

\section{Implications for the onset of radiation domination}\label{sec:implication}

\begin{figure}
    \centering
 \includegraphics[scale=0.6]{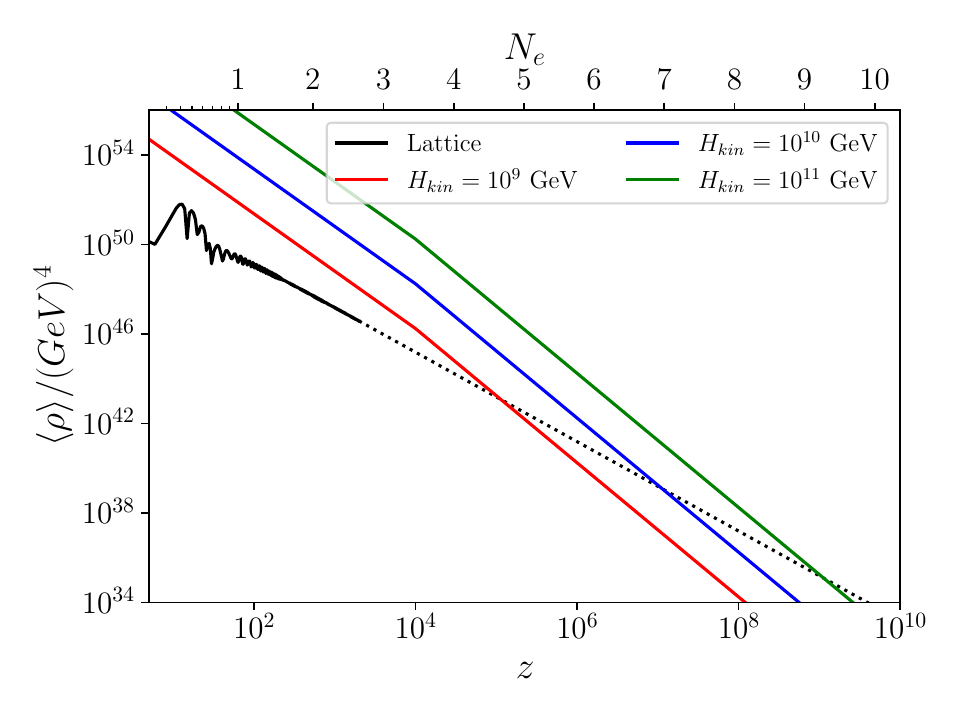}
    \caption{Comparison between the background energy density $\rho_B(z)$ in Eq.~\eqref{Benergy} for different $H_{\rm kin}$ scales and that transferred to the spectator field via the Hubble-induced symmetry breaking for the benchmark point  $\nu=10$, $\lambda=10^{-4}$, with $N_e$ the number of $e$-folds of kination. 
    \label{fig:energytotal}}
\end{figure}

The existence of a kination dominated era in quintessential inflation scenarios has strong implications for the onset of radiation domination \cite{Felder:1999pv,Rubio:2017gty}.  
In particular, even if the energy density transferred to the spectator field,
\begin{equation}\label{eq:ratio}
\rho_\chi(z) = \mathcal{X}\,\rho_Y(z)   = 16\nu^4  H_{\rm kin}^4  \, \left(\frac{a_{\rm kin}}{a(z)}\right)^4 \rho_Y(z) \,, 
\end{equation}
is initially  very  small, it  will  eventually become the dominant energy component due to the rapid decrease  of  the  background energy  density during this period, 
\begin{equation}\label{Benergy}
\rho_B(z)= 3M_P^2 H^2(z)\equiv \mathcal{X} \rho^{(r)}_B(z) \,  \,, \hspace{10mm} \rho^{(r)}_B(z) = \frac{ 3}{16\nu^4}  \left(\frac{M_P}{H_{\rm kin}}\right)^2 \left(\frac{a_{\rm kin}}{a(z)}\right)^2\,.
\end{equation}
The comparison between these two quantities for the benchmark point $\nu = 10$, $\lambda = 10^{-4}$
is displayed in Fig.~\ref{fig:energytotal} for different values of the Hubble rate at the onset of kinetic domination. As shown there, even for tiny initial ratios $\rho_\chi(z_i)/\rho_B(z_i)\sim {\cal O}(10^{-9})$, the Universe enters radiation domination in less than $10$ $e$-folds of kination.

It is important to emphasize at this point that the onset of radiation domination is \textit{not} related to the shape of the potential, as one could naively expect by considering the homogeneous approximation \cite{Opferkuch:2019zbd}, but rather to the late-time hierarchy $\braket{V}\ll \braket{K},\braket{G}$ observed in Fig.~\ref{fig:energies}. Indeed, for any virialized late-time scenario, [$1/2\langle{\dot{\chi}^2}\rangle=1/2\langle{(\nabla\chi/a)^2}\rangle+n\langle{V}\rangle $], involving a monomial potential $V\propto\chi^{2n}$ with $n\geq 1$ and negligible Hubble-induced interactions, one has \cite{Lozanov:2016hid,Lozanov:2017hjm} 
\begin{equation}
w_\chi=\frac{1}{3}+\frac{2}{3}\frac{(n-2)}{(n+1)+\langle(\nabla\chi/a)^2\rangle/\langle V\rangle}\,,
\end{equation}
and therefore $w_\chi\to 1/3$ if $ \langle(\nabla\chi/a)^2\rangle \gg \langle V\rangle$, in clear contrast with the value $w_\chi\to (n-1)/(n+1)$ obtained in the homogeneous limit  $\langle(\nabla\chi/a)^2\rangle \ll \langle V\rangle$ \cite{Turner:1983he,Johnson:2008se}.
 This can be explicitly verified by performing, for instance, a lattice simulation for a sextic potential $\frac{\sigma}{6\Lambda^2} \chi^6$, with $\sigma$ a coupling constant and $\Lambda$ a cutoff scale assumed to exceed $H_{\rm kin}$. In terms of the rescaled variables \eqref{eq:newvar}, the corresponding Klein-Gordon equation and potential energy density take the form 
\begin{equation}\label{EDOY6}
Y''-\nabla^2\, Y-M^2(z) Y+  \sigma_{\rm eff} \,  \,{\cal H}(z)\,  Y^5 =0\,, \hspace{10mm} V\equiv  \frac16\,  \sigma_{\rm eff} \,  \,{\cal H}(z)\, Y^6 \,,
\end{equation}
with 
\begin{equation}
\sigma_{\rm eff} \equiv \sigma  (2\nu)^3 \left(\frac{ H_{\rm kin}}{\Lambda}\right)^2
\end{equation}
an effective dimensionless coupling constant. 
\begin{figure}
    \centering
    \includegraphics[scale=0.45]{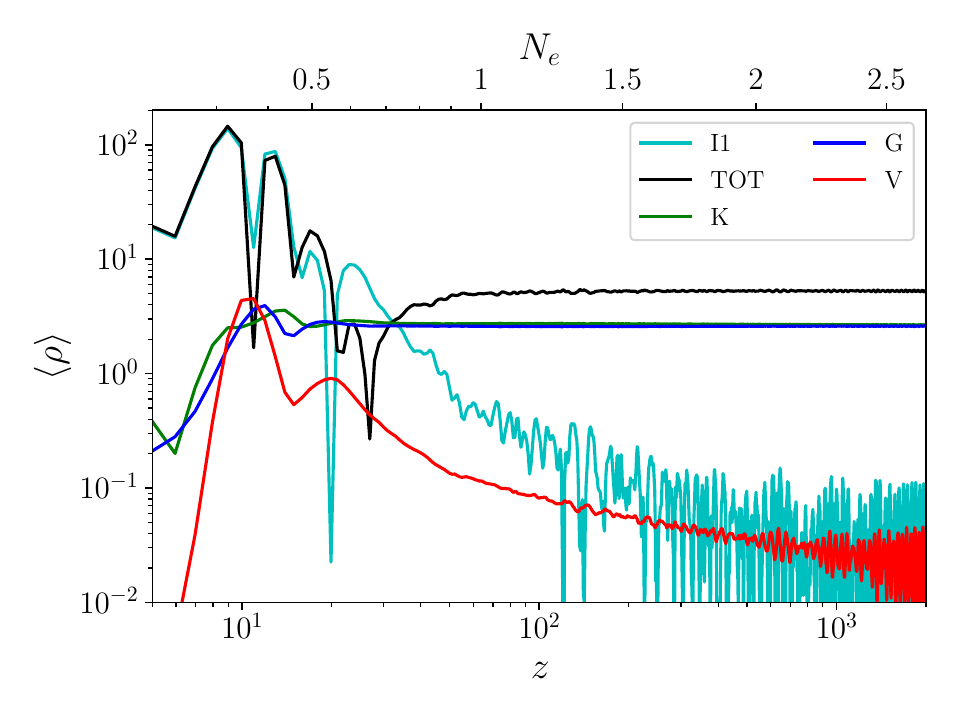}
    \includegraphics[scale=0.45]{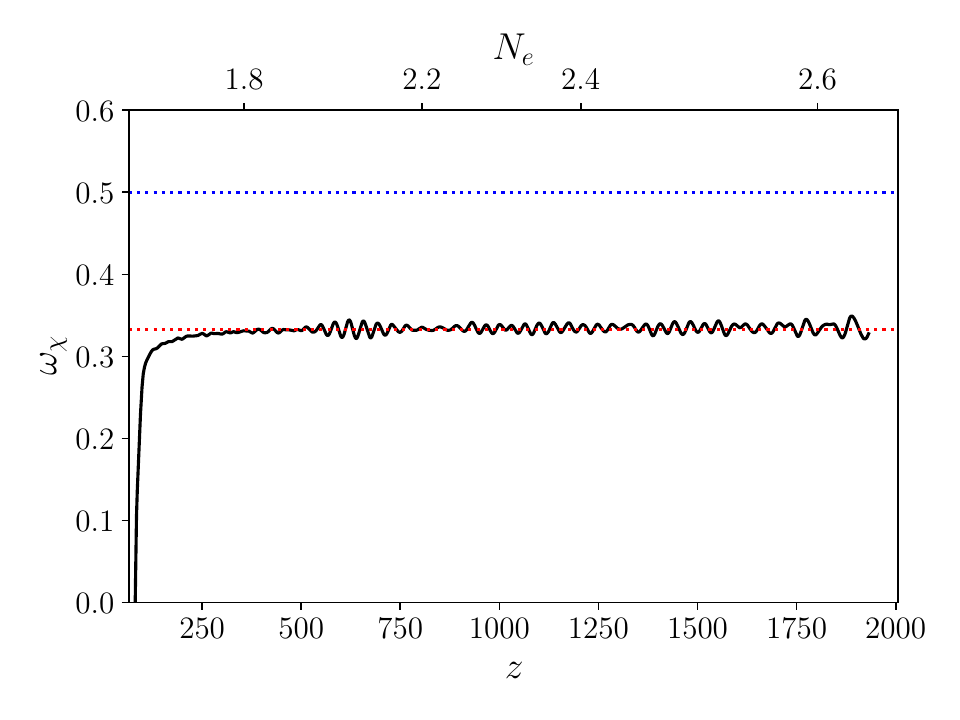}
   \caption{Energy distribution (left) and volume- and time-averaged equation of state \eqref{eq:wchi} (right) for the sextic scenario  in \eqref{EDOY6}. The asymptotic behaviour of $w_\chi$ differs clearly from the homogeneous expectation $w_\chi=1/2$, approaching a radiation-like value $w_\chi=1/3$. This illustrates the importance of accounting for the production of IR perturbations and their subsequent UV cascade when studying Ricci reheating scenarios.}
    \label{fig:eq_state_sextic}
\end{figure}
The evolution of the corresponding equation-of-state parameter $w_\chi$ for fiducial values $\nu=10$, $\sigma = 0.01$ and $\Lambda = 100\, H_{\rm kin}$ ($\sigma_{\rm eff}= 0.008$) is displayed in Fig.~\ref{fig:eq_state_sextic}. As expected, this quantity approaches a radiation-like value $w_\chi= 1/3$  rather than the $w_\chi= 1/2$ one following from the homogeneous approximation. Combined with kination, this observation allows for the generic onset of radiation domination for \textit{general interacting potentials}, significantly extending the results of Ref.~\cite{Opferkuch:2019zbd}.

\section{Assumptions and future research lines }\label{sec:limitations}

We have performed several approximations to reduce the physical scenario to a baseline model that could be treated in a simple way with analytical and numerical techniques. Relaxing any of these assumptions opens new avenues for model building.
\begin{enumerate}
\item \textit{Symmetry group:} We have assumed the spectator field to display a $Z_2$ symmetry, understanding this choice as a mere proof of concept of a much more general setting. In particular, we expect our results regarding the impact of perturbations on symmetry breaking and virialization to be extensible to more involved symmetry groups, provided that these are global.~\footnote{In these multifield settings an equipartition among the different field components is expected \cite{Felder:2000hr,Podolsky:2005bw,Easther:2006gt,Giblin:2014gra,Repond:2016sol}.} The extension to local symmetries could, on the other hand, lead to new phenomenology, certainly worthy to explore.
    \item \textit{Expansion of the Universe:} We have assumed the inflation-to-kination transition to be instantaneous as compared to the  temporal evolution of the spectator field. In practice, this corresponds to a quench approximation 
    \begin{equation}
    R(t<t_{\rm kin})>0 \,,     \hspace{10mm} R(t_{\rm kin}<t<t_r)<0\,,
    \end{equation}
    with $t_r$ the symmetry restoration time. On top of that, we assumed the spectator energy density to be subdominant during kination, such that the Ricci scalar can be well approximated by an homogeneous function of time depending only on the inflaton equation of state. While this is certainly a reasonable assumption for the main stages described in this paper, it should be eventually taken into account for a precise determination of the onset of radiation domination. 
    \item \textit{Large $\nu$ approximation:} 
    In order to minimize the impact of horizon effects, we have focused on a large $\nu$ limit leading to a reasonable separation between  ${\cal H}(z)$ and the typical momentum of the spectator field fluctuations, $\kappa_*(z)\equiv 2\sqrt{\nu+1}\, {\cal H}(z)$. This makes the integrated contribution of superhorizon modes reentering the horizon less and less important as compared with that of subhorizon scales never amplified by the inflationary dynamics, justifying the choice of a vacuum initial state with zero occupation numbers. We emphasize, however, that the asymptotic virialization observed in our simulations possesses an universal character, as we have explicitly verified by considering alternative initial conditions.
    
   \item \textit{Parameter space:} 
  We have restricted ourselves to model parameters leading to a maximum $10\%$ correction to the reduced Planck mass, i.e. $\xi \chi_{\rm max}^2/M_P^2\leq 0.1$ \cite{Bettoni:2019dcw}. Combined with the above two assumptions, this allows us to plausibly ignore metric fluctuations. Even within this self-consistency regime, our investigations are still limited to a small portion of the parameter space. Therefore, we cannot a priori exclude the existence of specific combinations of $\nu$ and $\lambda$ leading to different dynamics. We emphasize, however, that, given the decaying nature of the non-minimal coupling to gravity, the asymptotic behavior described in this paper is expected to be robust against variations of the model parameters. What could be affected at most is the duration of the tachyonic phase and the amount of defects that can be generated in the intermediate stages.
\item\textit{Couplings}: 
We have assumed the spectator field to be decoupled or weakly coupled to an unspecified inflationary sector. Unless forbidden by any symmetry, this requires some level of fine-tuning. On top of that, we have ignored potential couplings to other matter components that must necessarily exist to recover the Standard Model content and other relevant species as dark matter \cite{Bernal:2020bfj}, implicitly neglecting potential depletion and \textit{combined preheating} effects \cite{GarciaBellido:2008ab,Rubio:2015zia,Repond:2016sol,Fan:2021otj} that could  modify the approach to equilibrium. 

An interesting possibility along this line of thought is the potential identification of the spectator field with the Standard Model Higgs, as first suggested in Ref.~\cite{Figueroa:2016dsc} within the context of kinetic domination. This association would lead to a rather minimalistic heating scenario not involving additional degrees of freedom beyond the Standard Model content. One of the recurrent arguments for restricting the parameter space of this appealing possibility \cite{Figueroa:2016dsc,Opferkuch:2019zbd} is the apparent incompatibility of particle physics data with the absolute  stability of the Standard Model vacuum below the Planck scale \cite{Bezrukov:2012sa,Degrassi:2012ry}. We note, however, that the analyses of Ref.~\cite{Figueroa:2016dsc,Opferkuch:2019zbd} disregard implicitly the ambiguous relation between the \textit{Montecarlo reconstructed top quark mass} and the actual \textit{top pole mass}/Yukawa coupling entering in the renormalization group equations. As argued in Ref.~\cite{Bezrukov:2014ina}, once this source of uncertainty is taken into account, it is perfectly possible for the Standard Model vacuum to be absolutely stable all the way up till the Planck scale.  Indeed, the recent LHC measurements of the \textit{top pole mass} point intriguingly towards this scenario \cite{Aad:2019mkw,Sirunyan:2019zvx}.

Given the absolute stability of the Standard Model vacuum, the depletion of the spectator Higgs field in our setting is expected to proceed along the lines of Refs.~\cite{Enqvist:2013kaa,Enqvist:2014tta,Figueroa:2015rqa,Enqvist:2015sua,Figueroa:2017slm}, with potential differences associated to the bimodal pattern in Fig.~\ref{fig:histograms}, not accounted for in these papers.  The thermalization of the resulting Standard Model species is estimated to be almost instantaneous \cite{Bezrukov:2014ipa,Figueroa:2016dsc,Arnold:2002zm,Kurkela:2011ti}
\end{enumerate}

\section{Conclusions}\label{sec:conclusions}

In this work, we have made a step forward towards the complete characterization of  symmetry breaking in the context of Hubble-induced phase transitions, setting the stage for getting more reliable predictions to be eventually confronted with data. As a case of study, we investigated the full  symmetry-breaking dynamics of a $Z_2$ spectator field non-minimally coupled to gravity. The transition from the unbroken to the broken phase is induced by a change of sign in the Ricci scalar as the Universe evolves from inflation to an era in which the equation-of-state parameter is stiff. In order to follow the non-linear evolution of the spectator field until the development of an equilibrium configuration, we made use of classical lattice simulations, being the corresponding initial conditions analytically obtained from a complete quantum field theory description of the early dynamics.   

Our simulations allowed us to identify  
three different stages on the Hubble-induced symmetry breaking process. First, when the background acquires a stiff equation-of-state parameter immediately after the end of inflation, it induces a tachyonic instability for the spectator field fluctuations, which become amplified on infrared scales. This translates into the formation of rather homogeneous regions in real space separated by domain wall configurations. Second, as the field distribution splits and goes beyond the fading away minima of the effective potential, there appears a bi-modal oscillatory pattern that stays in place for a large number of oscillations. This stage is initially dominated by the non-minimal coupling to gravity, which become weaker and weaker as the Universe expands.  Finally, we observe a long virialization stage, where the total energy density becomes democratically distributed among the kinetic and gradient components and the spectrum develops a turbulent cascade towards the UV. The equation-of-state of the spectator field approaches then a radiation-like value $w_\chi=1/3$, independently of the shape of the potential. The memory loss induced by this asymptotic stage does not imply that all the traces of the temporary Hubble-induced symmetry breaking are completely erased. Among other consequences, the temporal existence of topological defects may lead to the production of a detectable gravitational waves' backgrounds \cite{Kamada:2014qja,Kamada:2015iga,Bettoni:2018pbl} or baryogenesis \cite{Bettoni:2018utf}, depending on the details of the symmetry breaking potential. We postpone the lattice characterization of these crucial aspects to a further publication.

\section*{Acknowledgments}
We thank Mark Hindmarsh for useful discussions on the physics and simulation of topological defects.
DB acknowledges support from the ``Atracción  del  Talento  Científico'' en Salamanca programme, from  Junta de Castilla y León and Fondo Europeo de desarrollo regional through the project SA0096P20 and from project PGC2018-096038-B-I00 by Spanish Ministerio de Ciencia, Innovación y Universidades.   JR acknowledges the support of the Fundação para a Ciência e a Tecnologia (Portugal) through the CEECIND/01091/2018 grant and Project~No.~UIDB/00099/2020. ALE (ORCID ID 0000-0002-1696-3579) is supported by the National Science Foundation grant PHY-1820872. This work has been possible thanks to the computational resources on the i2Basque academic network. The authors acknowledge the Tufts University High Performance Compute Cluster (https://it.tufts.edu/high-performance-computing) which was also utilized for the research reported in this paper.
\appendix

\section{The homogeneous approximation}\label{sec:appendix1} 

For the sake of completeness and in order facilitate the comparison with previous studies in the literature \cite{Figueroa:2016dsc,Dimopoulos:2018wfg,Opferkuch:2019zbd}, we discuss here the evolution of the zero mode in the absence of quantum fluctuations. In this classical and homogeneous limit, the solution of the equation of motion \eqref{EDOY} can be analytically obtained for both the tachyonic and pure quartic regimes. Assuming initial conditions $Y_{\rm kin}$ and $dY/d z\vert_{0} =Y_{\rm kin}/\nu $ and matching the solutions at the time 
\begin{equation}\label{eq:z_breaking}
\frac{z_{\rm b}}{\nu}=\left(\frac{8}{\lambda Y_{\rm kin}^2}\frac{2\nu-1}{2\nu+1}\right)^{1/(2\nu+3)}-1\,,
\end{equation} 
at which the quadratic and quartic terms in the $Y$-field potential become equal, we get \cite{Bettoni:2019dcw}
\begin{equation}\label{Yapprox}
Y(z)\simeq \begin{cases}
\frac{2\nu+1}{4\nu}Y_{\rm kin}\left(1+\frac{z}{\nu}\right)^{\nu+1/2} \hspace{24mm} {\rm for} \hspace{5mm} z\leq z_b\,, \\
Y_{\rm osc}\, \cn\left[\sqrt{\lambda}\,Y_{\rm osc} \,( z- z_{\rm osc}),\frac{1}{\sqrt{2}}\right]  \hspace{12mm} {\rm for} \hspace{5mm} z>z_b\,,
\end{cases}    
\end{equation}
where in the first expression we have neglected a subdominant decaying mode.  Here $\cn$ stands for the Jacobi elliptic cosine,
\begin{eqnarray}\label{Ymax}
Y_{\rm osc} =  \frac{Y(z_{b})}{\sqrt{2}}\left[1 + \left(1+\frac{2}{\beta^2}\frac{2\nu+1}{2\nu-1}\right)^{1/2}\right]^{1/2}
\end{eqnarray}
denotes the amplitude of the field at the time 
\begin{eqnarray}\label{zmax}
z_{\rm osc}=z_b+\frac{1}{\,\beta\sqrt{\lambda} Y_{\rm osc}}\arccos\left(\frac{Y(z_b)}{Y_{\rm osc}}\right)
\end{eqnarray}
at which the oscillatory part reaches its maximum ($z_{\rm osc}>z_b$) and $\beta=2\pi/T_x=0.8472$ is a numerical factor associated with the leading frequency of oscillation in the elliptic cosine series \cite{Bettoni:2019dcw}.

Although the solution \eqref{Yapprox} will break down as soon as gradients start to develop, it can be understood a relatively good description of the complete scenario at very early times, when all field fluctuations are still very close to the maximum of the effective potential.  In particular, we can use it to estimate the value of the spectator field equation-of-state parameter \eqref{eq:wchi} at the onset of kinetic domination.  Taking into account Eqs.~\eqref{eq:EEPPcomp} and \eqref{rhopY} at zero gradients, we get\footnote{If the decaying mode for the tachyonic regime in \eqref{Yapprox} is not omitted, we rather get
\begin{equation} \nonumber
    w_{\chi,{\rm ini}} = \frac{3 +12 \nu^2-32\nu^4}{3+36\nu^2}\,,
\end{equation}
which for the benchmark value $\nu=10$ translates into $w_{\chi,{\rm ini}}=-88.5$. Note, however, that the  growing mode becomes completely dominant already before the time $z_i$ at which we start our lattice simulations.
} 
\begin{equation}\label{wchiinit}
    w_{\chi, {\rm ini}} = 1-\frac{4\nu}{3}\,,
\end{equation}
which translates into a value $w_{ \chi, {\rm ini}} = -12.3$ for $\nu=10$,  in reasonable agreement with our lattice results and Ref.~\cite{Opferkuch:2019zbd}. 

\bibliographystyle{JHEP.bst}
\bibliography{HIPT}

\providecommand{\href}[2]{#2}\begingroup\raggedright\begin{thebibliography}{10}

\bibitem{Cormier:1998nt}
D.~Cormier and R.~Holman, \emph{{Spinodal inflation}},
  \href{https://doi.org/10.1103/PhysRevD.60.041301}{\emph{Phys. Rev. D}
  {\bfseries 60} (1999) 041301},
  [\href{https://arxiv.org/abs/hep-ph/9812476}{{\ttfamily hep-ph/9812476}}].

\bibitem{Tsujikawa:1999ni}
S.~Tsujikawa and T.~Torii, \emph{{Spinodal effect in the natural inflation
  model}}, \href{https://doi.org/10.1103/PhysRevD.62.043505}{\emph{Phys. Rev.
  D} {\bfseries 62} (2000) 043505},
  [\href{https://arxiv.org/abs/hep-ph/9912499}{{\ttfamily hep-ph/9912499}}].

\bibitem{Rubio:2019ypq}
J.~Rubio and E.~S. Tomberg, \emph{{Preheating in Palatini Higgs inflation}},
  \href{https://doi.org/10.1088/1475-7516/2019/04/021}{\emph{JCAP} {\bfseries
  04} (2019) 021}, [\href{https://arxiv.org/abs/1902.10148}{{\ttfamily
  1902.10148}}].

\bibitem{Karam:2021sno}
A.~Karam, E.~Tomberg and H.~Veerm\"ae, \emph{{Tachyonic Preheating in Palatini
  $R^2$ Inflation}},  \href{https://arxiv.org/abs/2102.02712}{{\ttfamily
  2102.02712}}.

\bibitem{Dufaux:2006ee}
J.~F. Dufaux, G.~N. Felder, L.~Kofman, M.~Peloso and D.~Podolsky,
  \emph{{Preheating with trilinear interactions: Tachyonic resonance}},
  \href{https://doi.org/10.1088/1475-7516/2006/07/006}{\emph{JCAP} {\bfseries
  0607} (2006) 006}, [\href{https://arxiv.org/abs/hep-ph/0602144}{{\ttfamily
  hep-ph/0602144}}].

\bibitem{Abolhasani:2009nb}
A.~A. Abolhasani, H.~Firouzjahi and M.~M. Sheikh-Jabbari, \emph{{Tachyonic
  Resonance Preheating in Expanding Universe}},
  \href{https://doi.org/10.1103/PhysRevD.81.043524}{\emph{Phys. Rev. D}
  {\bfseries 81} (2010) 043524},
  [\href{https://arxiv.org/abs/0912.1021}{{\ttfamily 0912.1021}}].

\bibitem{Felder:2000hj}
G.~N. Felder, J.~Garcia-Bellido, P.~B. Greene, L.~Kofman, A.~D. Linde and
  I.~Tkachev, \emph{{Dynamics of symmetry breaking and tachyonic preheating}},
  \href{https://doi.org/10.1103/PhysRevLett.87.011601}{\emph{Phys. Rev. Lett.}
  {\bfseries 87} (2001) 011601},
  [\href{https://arxiv.org/abs/hep-ph/0012142}{{\ttfamily hep-ph/0012142}}].

\bibitem{Felder:2001kt}
G.~N. Felder, L.~Kofman and A.~D. Linde, \emph{{Tachyonic instability and
  dynamics of spontaneous symmetry breaking}},
  \href{https://doi.org/10.1103/PhysRevD.64.123517}{\emph{Phys. Rev.}
  {\bfseries D64} (2001) 123517},
  [\href{https://arxiv.org/abs/hep-th/0106179}{{\ttfamily hep-th/0106179}}].

\bibitem{Copeland:2002ku}
E.~J. Copeland, S.~Pascoli and A.~Rajantie, \emph{{Dynamics of tachyonic
  preheating after hybrid inflation}},
  \href{https://doi.org/10.1103/PhysRevD.65.103517}{\emph{Phys. Rev.}
  {\bfseries D65} (2002) 103517},
  [\href{https://arxiv.org/abs/hep-ph/0202031}{{\ttfamily hep-ph/0202031}}].

\bibitem{GarciaBellido:2002aj}
J.~Garcia-Bellido, M.~Garcia~Perez and A.~Gonzalez-Arroyo, \emph{{Symmetry
  breaking and false vacuum decay after hybrid inflation}},
  \href{https://doi.org/10.1103/PhysRevD.67.103501}{\emph{Phys. Rev.}
  {\bfseries D67} (2003) 103501},
  [\href{https://arxiv.org/abs/hep-ph/0208228}{{\ttfamily hep-ph/0208228}}].

\bibitem{Giblin:2011yh}
J.~T. Giblin, Jr., L.~R. Price, X.~Siemens and B.~Vlcek, \emph{{Gravitational
  Waves from Global Second Order Phase Transitions}},
  \href{https://doi.org/10.1088/1475-7516/2012/11/006}{\emph{JCAP} {\bfseries
  11} (2012) 006}, [\href{https://arxiv.org/abs/1111.4014}{{\ttfamily
  1111.4014}}].

\bibitem{Bettoni:2018utf}
D.~Bettoni and J.~Rubio, \emph{{Quintessential Affleck-Dine baryogenesis with
  non-minimal couplings}},
  \href{https://doi.org/10.1016/j.physletb.2018.07.046}{\emph{Phys. Lett.}
  {\bfseries B784} (2018) 122--129},
  [\href{https://arxiv.org/abs/1805.02669}{{\ttfamily 1805.02669}}].

\bibitem{Bettoni:2018pbl}
D.~Bettoni, G.~Dom\'enech and J.~Rubio, \emph{{Gravitational waves from global
  cosmic strings in quintessential inflation}},
  \href{https://doi.org/10.1088/1475-7516/2019/02/034}{\emph{JCAP} {\bfseries
  1902} (2019) 034}, [\href{https://arxiv.org/abs/1810.11117}{{\ttfamily
  1810.11117}}].

\bibitem{Bettoni:2019dcw}
D.~Bettoni and J.~Rubio, \emph{{Hubble-induced phase transitions: Walls are not
  forever}}, \href{https://doi.org/10.1088/1475-7516/2020/01/002}{\emph{JCAP}
  {\bfseries 2001} (2020) 002},
  [\href{https://arxiv.org/abs/1911.03484}{{\ttfamily 1911.03484}}].

\bibitem{Figueroa:2016dsc}
D.~G. Figueroa and C.~T. Byrnes, \emph{{The Standard Model Higgs as the origin
  of the hot Big Bang}},
  \href{https://doi.org/10.1016/j.physletb.2017.01.059}{\emph{Phys. Lett.}
  {\bfseries B767} (2017) 272--277},
  [\href{https://arxiv.org/abs/1604.03905}{{\ttfamily 1604.03905}}].

\bibitem{Dimopoulos:2018wfg}
K.~Dimopoulos and T.~Markkanen, \emph{{Non-minimal gravitational reheating
  during kination}},
  \href{https://doi.org/10.1088/1475-7516/2018/06/021}{\emph{JCAP} {\bfseries
  1806} (2018) 021}, [\href{https://arxiv.org/abs/1803.07399}{{\ttfamily
  1803.07399}}].

\bibitem{Opferkuch:2019zbd}
T.~Opferkuch, P.~Schwaller and B.~A. Stefanek, \emph{{Ricci Reheating}},
  \href{https://doi.org/10.1088/1475-7516/2019/07/016}{\emph{JCAP} {\bfseries
  1907} (2019) 016}, [\href{https://arxiv.org/abs/1905.06823}{{\ttfamily
  1905.06823}}].

\bibitem{Fairbairn:2018bsw}
M.~Fairbairn, K.~Kainulainen, T.~Markkanen and S.~Nurmi, \emph{{Despicable Dark
  Relics: generated by gravity with unconstrained masses}},
  \href{https://doi.org/10.1088/1475-7516/2019/04/005}{\emph{JCAP} {\bfseries
  1904} (2019) 005}, [\href{https://arxiv.org/abs/1808.08236}{{\ttfamily
  1808.08236}}].

\bibitem{Laulumaa:2020pqi}
L.~Laulumaa, T.~Markkanen and S.~Nurmi, \emph{{Primordial dark matter from
  curvature induced symmetry breaking}},
  \href{https://doi.org/10.1088/1475-7516/2020/08/002}{\emph{JCAP} {\bfseries
  08} (2020) 002}, [\href{https://arxiv.org/abs/2005.04061}{{\ttfamily
  2005.04061}}].

\bibitem{Allahverdi:2020bys}
R.~Allahverdi et~al., \emph{{The First Three Seconds: a Review of Possible
  Expansion Histories of the Early Universe}},
  \href{https://arxiv.org/abs/2006.16182}{{\ttfamily 2006.16182}}.

\bibitem{Nguyen:2019kbm}
R.~Nguyen, J.~van~de Vis, E.~I. Sfakianakis, J.~T. Giblin and D.~I. Kaiser,
  \emph{{Nonlinear Dynamics of Preheating after Multifield Inflation with
  Nonminimal Couplings}},
  \href{https://doi.org/10.1103/PhysRevLett.123.171301}{\emph{Phys. Rev. Lett.}
  {\bfseries 123} (2019) 171301},
  [\href{https://arxiv.org/abs/1905.12562}{{\ttfamily 1905.12562}}].

\bibitem{vandeVis:2020qcp}
J.~van~de Vis, R.~Nguyen, E.~I. Sfakianakis, J.~T. Giblin and D.~I. Kaiser,
  \emph{{Time scales for nonlinear processes in preheating after multifield
  inflation with nonminimal couplings}},
  \href{https://doi.org/10.1103/PhysRevD.102.043528}{\emph{Phys. Rev. D}
  {\bfseries 102} (2020) 043528},
  [\href{https://arxiv.org/abs/2005.00433}{{\ttfamily 2005.00433}}].

\bibitem{Khlebnikov:1996mc}
S.~Y. Khlebnikov and I.~I. Tkachev, \emph{{Classical decay of inflaton}},
  \href{https://doi.org/10.1103/PhysRevLett.77.219}{\emph{Phys. Rev. Lett.}
  {\bfseries 77} (1996) 219--222},
  [\href{https://arxiv.org/abs/hep-ph/9603378}{{\ttfamily hep-ph/9603378}}].

\bibitem{Micha:2004bv}
R.~Micha and I.~I. Tkachev, \emph{{Turbulent thermalization}},
  \href{https://doi.org/10.1103/PhysRevD.70.043538}{\emph{Phys. Rev.}
  {\bfseries D70} (2004) 043538},
  [\href{https://arxiv.org/abs/hep-ph/0403101}{{\ttfamily hep-ph/0403101}}].

\bibitem{Micha:2002ey}
R.~Micha and I.~I. Tkachev, \emph{{Relativistic turbulence: A Long way from
  preheating to equilibrium}},
  \href{https://doi.org/10.1103/PhysRevLett.90.121301}{\emph{Phys. Rev. Lett.}
  {\bfseries 90} (2003) 121301},
  [\href{https://arxiv.org/abs/hep-ph/0210202}{{\ttfamily hep-ph/0210202}}].

\bibitem{Berges:2010ez}
J.~Berges and D.~Sexty, \emph{{Strong versus weak wave-turbulence in
  relativistic field theory}},
  \href{https://doi.org/10.1103/PhysRevD.83.085004}{\emph{Phys. Rev. D}
  {\bfseries 83} (2011) 085004},
  [\href{https://arxiv.org/abs/1012.5944}{{\ttfamily 1012.5944}}].

\bibitem{Wetterich:1987fm}
C.~Wetterich, \emph{{Cosmology and the Fate of Dilatation Symmetry}},
  \href{https://doi.org/10.1016/0550-3213(88)90193-9}{\emph{Nucl. Phys.}
  {\bfseries B302} (1988) 668--696},
  [\href{https://arxiv.org/abs/1711.03844}{{\ttfamily 1711.03844}}].

\bibitem{Wetterich:1994bg}
C.~Wetterich, \emph{{The Cosmon model for an asymptotically vanishing time
  dependent cosmological 'constant'}}, {\emph{Astron. Astrophys.} {\bfseries
  301} (1995) 321--328},
  [\href{https://arxiv.org/abs/hep-th/9408025}{{\ttfamily hep-th/9408025}}].

\bibitem{Peebles:1998qn}
P.~J.~E. Peebles and A.~Vilenkin, \emph{{Quintessential inflation}},
  \href{https://doi.org/10.1103/PhysRevD.59.063505}{\emph{Phys. Rev.}
  {\bfseries D59} (1999) 063505},
  [\href{https://arxiv.org/abs/astro-ph/9810509}{{\ttfamily
  astro-ph/9810509}}].

\bibitem{Spokoiny:1993kt}
B.~Spokoiny, \emph{{Deflationary universe scenario}},
  \href{https://doi.org/10.1016/0370-2693(93)90155-B}{\emph{Phys. Lett.}
  {\bfseries B315} (1993) 40--45},
  [\href{https://arxiv.org/abs/gr-qc/9306008}{{\ttfamily gr-qc/9306008}}].

\bibitem{Brax:2005uf}
P.~Brax and J.~Martin, \emph{{Coupling quintessence to inflation in
  supergravity}}, \href{https://doi.org/10.1103/PhysRevD.71.063530}{\emph{Phys.
  Rev.} {\bfseries D71} (2005) 063530},
  [\href{https://arxiv.org/abs/astro-ph/0502069}{{\ttfamily
  astro-ph/0502069}}].

\bibitem{Wetterich:2013jsa}
C.~Wetterich, \emph{{Variable gravity Universe}},
  \href{https://doi.org/10.1103/PhysRevD.89.024005}{\emph{Phys. Rev.}
  {\bfseries D89} (2014) 024005},
  [\href{https://arxiv.org/abs/1308.1019}{{\ttfamily 1308.1019}}].

\bibitem{Wetterich:2014gaa}
C.~Wetterich, \emph{{Inflation, quintessence, and the origin of mass}},
  \href{https://doi.org/10.1016/j.nuclphysb.2015.05.019}{\emph{Nucl. Phys.}
  {\bfseries B897} (2015) 111--178},
  [\href{https://arxiv.org/abs/1408.0156}{{\ttfamily 1408.0156}}].

\bibitem{Hossain:2014xha}
M.~W. Hossain, R.~Myrzakulov, M.~Sami and E.~N. Saridakis, \emph{{Variable
  gravity: A suitable framework for quintessential inflation}},
  \href{https://doi.org/10.1103/PhysRevD.90.023512}{\emph{Phys. Rev.}
  {\bfseries D90} (2014) 023512},
  [\href{https://arxiv.org/abs/1402.6661}{{\ttfamily 1402.6661}}].

\bibitem{Agarwal:2017wxo}
A.~Agarwal, R.~Myrzakulov, M.~Sami and N.~K. Singh, \emph{{Quintessential
  inflation in a thawing realization}},
  \href{https://doi.org/10.1016/j.physletb.2017.04.066}{\emph{Phys. Lett.}
  {\bfseries B770} (2017) 200--208},
  [\href{https://arxiv.org/abs/1708.00156}{{\ttfamily 1708.00156}}].

\bibitem{Geng:2017mic}
C.-Q. Geng, C.-C. Lee, M.~Sami, E.~N. Saridakis and A.~A. Starobinsky,
  \emph{{Observational constraints on successful model of quintessential
  Inflation}}, \href{https://doi.org/10.1088/1475-7516/2017/06/011}{\emph{JCAP}
  {\bfseries 1706} (2017) 011},
  [\href{https://arxiv.org/abs/1705.01329}{{\ttfamily 1705.01329}}].

\bibitem{Dimopoulos:2017zvq}
K.~Dimopoulos and C.~Owen, \emph{{Quintessential Inflation with
  $\alpha$-attractors}},
  \href{https://doi.org/10.1088/1475-7516/2017/06/027}{\emph{JCAP} {\bfseries
  1706} (2017) 027}, [\href{https://arxiv.org/abs/1703.00305}{{\ttfamily
  1703.00305}}].

\bibitem{Rubio:2017gty}
J.~Rubio and C.~Wetterich, \emph{{Emergent scale symmetry: Connecting inflation
  and dark energy}},
  \href{https://doi.org/10.1103/PhysRevD.96.063509}{\emph{Phys. Rev.}
  {\bfseries D96} (2017) 063509},
  [\href{https://arxiv.org/abs/1705.00552}{{\ttfamily 1705.00552}}].

\bibitem{Dimopoulos:2017tud}
K.~Dimopoulos, L.~Donaldson~Wood and C.~Owen, \emph{{Instant preheating in
  quintessential inflation with $\alpha$-attractors}},
  \href{https://doi.org/10.1103/PhysRevD.97.063525}{\emph{Phys. Rev.}
  {\bfseries D97} (2018) 063525},
  [\href{https://arxiv.org/abs/1712.01760}{{\ttfamily 1712.01760}}].

\bibitem{Akrami:2017cir}
Y.~Akrami, R.~Kallosh, A.~Linde and V.~Vardanyan, \emph{{Dark energy,
  $\alpha$-attractors, and large-scale structure surveys}},
  \href{https://doi.org/10.1088/1475-7516/2018/06/041}{\emph{JCAP} {\bfseries
  1806} (2018) 041}, [\href{https://arxiv.org/abs/1712.09693}{{\ttfamily
  1712.09693}}].

\bibitem{Garcia-Garcia:2018hlc}
C.~García-García, E.~V. Linder, P.~Ruíz-Lapuente and M.~Zumalacárregui,
  \emph{{Dark energy from $\alpha$-attractors: phenomenology and observational
  constraints}},
  \href{https://doi.org/10.1088/1475-7516/2018/08/022}{\emph{JCAP} {\bfseries
  1808} (2018) 022}, [\href{https://arxiv.org/abs/1803.00661}{{\ttfamily
  1803.00661}}].

\bibitem{Markkanen:2017dlc}
T.~Markkanen, S.~Nurmi and A.~Rajantie, \emph{{Do metric fluctuations affect
  the Higgs dynamics during inflation?}},
  \href{https://doi.org/10.1088/1475-7516/2017/12/026}{\emph{JCAP} {\bfseries
  12} (2017) 026}, [\href{https://arxiv.org/abs/1707.00866}{{\ttfamily
  1707.00866}}].

\bibitem{Ford:1987de}
L.~H. Ford, \emph{{Cosmological constant damping by unstable scalar fields}},
  \href{https://doi.org/10.1103/PhysRevD.35.2339}{\emph{Phys. Rev. D}
  {\bfseries 35} (1987) 2339}.

\bibitem{Ford:2000xg}
L.~H. Ford and T.~A. Roman, \emph{{Classical scalar fields and violations of
  the second law}},
  \href{https://doi.org/10.1103/PhysRevD.64.024023}{\emph{Phys. Rev. D}
  {\bfseries 64} (2001) 024023},
  [\href{https://arxiv.org/abs/gr-qc/0009076}{{\ttfamily gr-qc/0009076}}].

\bibitem{Bekenstein:1975ww}
J.~D. Bekenstein, \emph{{Nonsingular General Relativistic Cosmologies}},
  \href{https://doi.org/10.1103/PhysRevD.11.2072}{\emph{Phys. Rev. D}
  {\bfseries 11} (1975) 2072--2075}.

\bibitem{Flanagan:1996gw}
E.~E. Flanagan and R.~M. Wald, \emph{{Does back reaction enforce the averaged
  null energy condition in semiclassical gravity?}},
  \href{https://doi.org/10.1103/PhysRevD.54.6233}{\emph{Phys. Rev. D}
  {\bfseries 54} (1996) 6233--6283},
  [\href{https://arxiv.org/abs/gr-qc/9602052}{{\ttfamily gr-qc/9602052}}].

\bibitem{Polarski:1995jg}
D.~Polarski and A.~A. Starobinsky, \emph{{Semiclassicality and decoherence of
  cosmological perturbations}},
  \href{https://doi.org/10.1088/0264-9381/13/3/006}{\emph{Class. Quant. Grav.}
  {\bfseries 13} (1996) 377--392},
  [\href{https://arxiv.org/abs/gr-qc/9504030}{{\ttfamily gr-qc/9504030}}].

\bibitem{Lesgourgues:1996jc}
J.~Lesgourgues, D.~Polarski and A.~A. Starobinsky, \emph{{Quantum to classical
  transition of cosmological perturbations for nonvacuum initial states}},
  \href{https://doi.org/10.1016/S0550-3213(97)00224-1}{\emph{Nucl. Phys.}
  {\bfseries B497} (1997) 479--510},
  [\href{https://arxiv.org/abs/gr-qc/9611019}{{\ttfamily gr-qc/9611019}}].

\bibitem{Kiefer:1998jk}
C.~Kiefer and D.~Polarski, \emph{{Emergence of classicality for primordial
  fluctuations: Concepts and analogies}},
  \href{https://doi.org/10.1002/andp.2090070302}{\emph{Annalen Phys.}
  {\bfseries 7} (1998) 137--158},
  [\href{https://arxiv.org/abs/gr-qc/9805014}{{\ttfamily gr-qc/9805014}}].

\bibitem{abramowitz+stegun}
M.~Abramowitz and I.~A. Stegun, \emph{Handbook of Mathematical Functions with
  Formulas, Graphs, and Mathematical Tables}.
\newblock Dover, New York, ninth dover printing, tenth gpo printing~ed., 1964.

\bibitem{Cornwall:1974vz}
J.~M. Cornwall, R.~Jackiw and E.~Tomboulis, \emph{{Effective Action for
  Composite Operators}},
  \href{https://doi.org/10.1103/PhysRevD.10.2428}{\emph{Phys. Rev. D}
  {\bfseries 10} (1974) 2428--2445}.

\bibitem{Berges:2000ur}
J.~Berges and J.~Cox, \emph{{Thermalization of quantum fields from time
  reversal invariant evolution equations}},
  \href{https://doi.org/10.1016/S0370-2693(01)01004-8}{\emph{Phys. Lett. B}
  {\bfseries 517} (2001) 369--374},
  [\href{https://arxiv.org/abs/hep-ph/0006160}{{\ttfamily hep-ph/0006160}}].

\bibitem{Arrizabalaga:2004iw}
A.~Arrizabalaga, J.~Smit and A.~Tranberg, \emph{{Tachyonic preheating using
  2PI-1/N dynamics and the classical approximation}},
  \href{https://doi.org/10.1088/1126-6708/2004/10/017}{\emph{JHEP} {\bfseries
  10} (2004) 017}, [\href{https://arxiv.org/abs/hep-ph/0409177}{{\ttfamily
  hep-ph/0409177}}].

\bibitem{Daverio:2015ryl}
D.~Daverio, M.~Hindmarsh and N.~Bevis, \emph{{Latfield2: A c++ library for
  classical lattice field theory}},
  \href{https://arxiv.org/abs/1508.05610}{{\ttfamily 1508.05610}}.

\bibitem{Daverio:2015nva}
D.~Daverio, M.~Hindmarsh, M.~Kunz, J.~Lizarraga and J.~Urrestilla,
  \emph{{Energy-momentum correlations for Abelian Higgs cosmic strings}},
  \href{https://doi.org/10.1103/PhysRevD.95.049903}{\emph{Phys. Rev. D}
  {\bfseries 93} (2016) 085014},
  [\href{https://arxiv.org/abs/1510.05006}{{\ttfamily 1510.05006}}].

\bibitem{Hindmarsh:2017qff}
M.~Hindmarsh, J.~Lizarraga, J.~Urrestilla, D.~Daverio and M.~Kunz,
  \emph{{Scaling from gauge and scalar radiation in Abelian Higgs string
  networks}}, \href{https://doi.org/10.1103/PhysRevD.96.023525}{\emph{Phys.
  Rev. D} {\bfseries 96} (2017) 023525},
  [\href{https://arxiv.org/abs/1703.06696}{{\ttfamily 1703.06696}}].

\bibitem{Lopez-Eiguren:2017dmc}
A.~Lopez-Eiguren, J.~Lizarraga, M.~Hindmarsh and J.~Urrestilla, \emph{{Cosmic
  Microwave Background constraints for global strings and global monopoles}},
  \href{https://doi.org/10.1088/1475-7516/2017/07/026}{\emph{JCAP} {\bfseries
  07} (2017) 026}, [\href{https://arxiv.org/abs/1705.04154}{{\ttfamily
  1705.04154}}].

\bibitem{Hindmarsh:2019csc}
M.~Hindmarsh, J.~Lizarraga, A.~Lopez-Eiguren and J.~Urrestilla, \emph{{Scaling
  Density of Axion Strings}},
  \href{https://doi.org/10.1103/PhysRevLett.124.021301}{\emph{Phys. Rev. Lett.}
  {\bfseries 124} (2020) 021301},
  [\href{https://arxiv.org/abs/1908.03522}{{\ttfamily 1908.03522}}].

\bibitem{Hindmarsh:2021vih}
M.~Hindmarsh, J.~Lizarraga, A.~Lopez-Eiguren and J.~Urrestilla, \emph{{Approach
  to scaling in axion string networks}},
  \href{https://doi.org/10.1103/PhysRevD.103.103534}{\emph{Phys. Rev. D}
  {\bfseries 103} (2021) 103534},
  [\href{https://arxiv.org/abs/2102.07723}{{\ttfamily 2102.07723}}].

\bibitem{Salle:2000hd}
M.~Salle, J.~Smit and J.~C. Vink, \emph{{Thermalization in a Hartree ensemble
  approximation to quantum field dynamics}},
  \href{https://doi.org/10.1103/PhysRevD.64.025016}{\emph{Phys. Rev. D}
  {\bfseries 64} (2001) 025016},
  [\href{https://arxiv.org/abs/hep-ph/0012346}{{\ttfamily hep-ph/0012346}}].

\bibitem{Salle:2000jb}
M.~Salle, J.~Smit and J.~C. Vink, \emph{{Staying thermal with Hartree ensemble
  approximations}},
  \href{https://doi.org/10.1016/S0550-3213(01)00659-9}{\emph{Nucl. Phys. B}
  {\bfseries 625} (2002) 495--511},
  [\href{https://arxiv.org/abs/hep-ph/0012362}{{\ttfamily hep-ph/0012362}}].

\bibitem{Berges:2013lsa}
J.~Berges, K.~Boguslavski, S.~Schlichting and R.~Venugopalan, \emph{{Basin of
  attraction for turbulent thermalization and the range of validity of
  classical-statistical simulations}},
  \href{https://doi.org/10.1007/JHEP05(2014)054}{\emph{JHEP} {\bfseries 05}
  (2014) 054}, [\href{https://arxiv.org/abs/1312.5216}{{\ttfamily 1312.5216}}].

\bibitem{Destri:2004ck}
C.~Destri and H.~J. de~Vega, \emph{{Ultraviolet cascade in the thermalization
  of the classical phi**4 theory in 3+1 dimensions}},
  \href{https://doi.org/10.1103/PhysRevD.73.025014}{\emph{Phys. Rev. D}
  {\bfseries 73} (2006) 025014},
  [\href{https://arxiv.org/abs/hep-ph/0410280}{{\ttfamily hep-ph/0410280}}].

\bibitem{Boyanovsky:2003tc}
D.~Boyanovsky, C.~Destri and H.~J. de~Vega, \emph{{The Approach to
  thermalization in the classical phi**4 theory in (1+1)-dimensions: Energy
  cascades and universal scaling}},
  \href{https://doi.org/10.1103/PhysRevD.69.045003}{\emph{Phys. Rev. D}
  {\bfseries 69} (2004) 045003},
  [\href{https://arxiv.org/abs/hep-ph/0306124}{{\ttfamily hep-ph/0306124}}].

\bibitem{Podolsky:2005bw}
D.~I. Podolsky, G.~N. Felder, L.~Kofman and M.~Peloso, \emph{{Equation of state
  and beginning of thermalization after preheating}},
  \href{https://doi.org/10.1103/PhysRevD.73.023501}{\emph{Phys. Rev. D}
  {\bfseries 73} (2006) 023501},
  [\href{https://arxiv.org/abs/hep-ph/0507096}{{\ttfamily hep-ph/0507096}}].

\bibitem{Felder:1999pv}
G.~N. Felder, L.~Kofman and A.~D. Linde, \emph{{Inflation and preheating in NO
  models}}, \href{https://doi.org/10.1103/PhysRevD.60.103505}{\emph{Phys. Rev.}
  {\bfseries D60} (1999) 103505},
  [\href{https://arxiv.org/abs/hep-ph/9903350}{{\ttfamily hep-ph/9903350}}].

\bibitem{Lozanov:2016hid}
K.~D. Lozanov and M.~A. Amin, \emph{{Equation of State and Duration to
  Radiation Domination after Inflation}},
  \href{https://doi.org/10.1103/PhysRevLett.119.061301}{\emph{Phys. Rev. Lett.}
  {\bfseries 119} (2017) 061301},
  [\href{https://arxiv.org/abs/1608.01213}{{\ttfamily 1608.01213}}].

\bibitem{Lozanov:2017hjm}
K.~D. Lozanov and M.~A. Amin, \emph{{Self-resonance after inflation: oscillons,
  transients and radiation domination}},
  \href{https://doi.org/10.1103/PhysRevD.97.023533}{\emph{Phys. Rev. D}
  {\bfseries 97} (2018) 023533},
  [\href{https://arxiv.org/abs/1710.06851}{{\ttfamily 1710.06851}}].

\bibitem{Turner:1983he}
M.~S. Turner, \emph{{Coherent Scalar Field Oscillations in an Expanding
  Universe}}, \href{https://doi.org/10.1103/PhysRevD.28.1243}{\emph{Phys. Rev.}
  {\bfseries D28} (1983) 1243}.

\bibitem{Johnson:2008se}
M.~C. Johnson and M.~Kamionkowski, \emph{{Dynamical and Gravitational
  Instability of Oscillating-Field Dark Energy and Dark Matter}},
  \href{https://doi.org/10.1103/PhysRevD.78.063010}{\emph{Phys. Rev. D}
  {\bfseries 78} (2008) 063010},
  [\href{https://arxiv.org/abs/0805.1748}{{\ttfamily 0805.1748}}].

\bibitem{Felder:2000hr}
G.~N. Felder and L.~Kofman, \emph{{The Development of equilibrium after
  preheating}}, \href{https://doi.org/10.1103/PhysRevD.63.103503}{\emph{Phys.
  Rev. D} {\bfseries 63} (2001) 103503},
  [\href{https://arxiv.org/abs/hep-ph/0011160}{{\ttfamily hep-ph/0011160}}].

\bibitem{Easther:2006gt}
R.~Easther and E.~A. Lim, \emph{{Stochastic gravitational wave production after
  inflation}}, \href{https://doi.org/10.1088/1475-7516/2006/04/010}{\emph{JCAP}
  {\bfseries 04} (2006) 010},
  [\href{https://arxiv.org/abs/astro-ph/0601617}{{\ttfamily
  astro-ph/0601617}}].

\bibitem{Giblin:2014gra}
J.~T. Giblin and E.~Thrane, \emph{{Estimates of maximum energy density of
  cosmological gravitational-wave backgrounds}},
  \href{https://doi.org/10.1103/PhysRevD.90.107502}{\emph{Phys. Rev. D}
  {\bfseries 90} (2014) 107502},
  [\href{https://arxiv.org/abs/1410.4779}{{\ttfamily 1410.4779}}].

\bibitem{Repond:2016sol}
J.~Repond and J.~Rubio, \emph{{Combined Preheating on the lattice with
  applications to Higgs inflation}},
  \href{https://doi.org/10.1088/1475-7516/2016/07/043}{\emph{JCAP} {\bfseries
  1607} (2016) 043}, [\href{https://arxiv.org/abs/1604.08238}{{\ttfamily
  1604.08238}}].

\bibitem{Bernal:2020bfj}
N.~Bernal, J.~Rubio and H.~Veerm\"ae, \emph{{Boosting Ultraviolet Freeze-in in
  NO Models}}, \href{https://doi.org/10.1088/1475-7516/2020/06/047}{\emph{JCAP}
  {\bfseries 06} (2020) 047},
  [\href{https://arxiv.org/abs/2004.13706}{{\ttfamily 2004.13706}}].

\bibitem{GarciaBellido:2008ab}
J.~Garcia-Bellido, D.~G. Figueroa and J.~Rubio, \emph{{Preheating in the
  Standard Model with the Higgs-Inflaton coupled to gravity}},
  \href{https://doi.org/10.1103/PhysRevD.79.063531}{\emph{Phys. Rev.}
  {\bfseries D79} (2009) 063531},
  [\href{https://arxiv.org/abs/0812.4624}{{\ttfamily 0812.4624}}].

\bibitem{Rubio:2015zia}
J.~Rubio, \emph{{Higgs inflation and vacuum stability}},
  \href{https://doi.org/10.1088/1742-6596/631/1/012032}{\emph{J. Phys. Conf.
  Ser.} {\bfseries 631} (2015) 012032},
  [\href{https://arxiv.org/abs/1502.07952}{{\ttfamily 1502.07952}}].

\bibitem{Fan:2021otj}
J.~Fan, K.~D. Lozanov and Q.~Lu, \emph{{Spillway Preheating}},
  \href{https://doi.org/10.1007/JHEP05(2021)069}{\emph{JHEP} {\bfseries 05}
  (2021) 069}, [\href{https://arxiv.org/abs/2101.11008}{{\ttfamily
  2101.11008}}].

\bibitem{Bezrukov:2012sa}
F.~Bezrukov, M.~Y. Kalmykov, B.~A. Kniehl and M.~Shaposhnikov, \emph{{Higgs
  Boson Mass and New Physics}},
  \href{https://doi.org/10.1007/JHEP10(2012)140}{\emph{JHEP} {\bfseries 10}
  (2012) 140}, [\href{https://arxiv.org/abs/1205.2893}{{\ttfamily 1205.2893}}].

\bibitem{Degrassi:2012ry}
G.~Degrassi, S.~Di~Vita, J.~Elias-Miro, J.~R. Espinosa, G.~F. Giudice,
  G.~Isidori et~al., \emph{{Higgs mass and vacuum stability in the Standard
  Model at NNLO}}, \href{https://doi.org/10.1007/JHEP08(2012)098}{\emph{JHEP}
  {\bfseries 08} (2012) 098},
  [\href{https://arxiv.org/abs/1205.6497}{{\ttfamily 1205.6497}}].

\bibitem{Bezrukov:2014ina}
F.~Bezrukov and M.~Shaposhnikov, \emph{{Why should we care about the top quark
  Yukawa coupling?}}, \href{https://doi.org/10.1134/S1063776115030152}{\emph{J.
  Exp. Theor. Phys.} {\bfseries 120} (2015) 335--343},
  [\href{https://arxiv.org/abs/1411.1923}{{\ttfamily 1411.1923}}].

\bibitem{Aad:2019mkw}
{\scshape ATLAS} collaboration, G.~Aad et~al., \emph{{Measurement of the
  top-quark mass in $t\bar{t}+1$-jet events collected with the ATLAS detector
  in $pp$ collisions at $\sqrt{s}=8$ TeV}},
  \href{https://doi.org/10.1007/JHEP11(2019)150}{\emph{JHEP} {\bfseries 11}
  (2019) 150}, [\href{https://arxiv.org/abs/1905.02302}{{\ttfamily
  1905.02302}}].

\bibitem{Sirunyan:2019zvx}
{\scshape CMS} collaboration, A.~M. Sirunyan et~al., \emph{{Measurement of
  $\mathrm{t\bar t}$ normalised multi-differential cross sections in pp
  collisions at $\sqrt s=13$ TeV, and simultaneous determination of the strong
  coupling strength, top quark pole mass, and parton distribution functions}},
  \href{https://doi.org/10.1140/epjc/s10052-020-7917-7}{\emph{Eur. Phys. J. C}
  {\bfseries 80} (2020) 658},
  [\href{https://arxiv.org/abs/1904.05237}{{\ttfamily 1904.05237}}].

\bibitem{Enqvist:2013kaa}
K.~Enqvist, T.~Meriniemi and S.~Nurmi, \emph{{Generation of the Higgs
  Condensate and Its Decay after Inflation}},
  \href{https://doi.org/10.1088/1475-7516/2013/10/057}{\emph{JCAP} {\bfseries
  10} (2013) 057}, [\href{https://arxiv.org/abs/1306.4511}{{\ttfamily
  1306.4511}}].

\bibitem{Enqvist:2014tta}
K.~Enqvist, S.~Nurmi and S.~Rusak, \emph{{Non-Abelian dynamics in the resonant
  decay of the Higgs after inflation}},
  \href{https://doi.org/10.1088/1475-7516/2014/10/064}{\emph{JCAP} {\bfseries
  10} (2014) 064}, [\href{https://arxiv.org/abs/1404.3631}{{\ttfamily
  1404.3631}}].

\bibitem{Figueroa:2015rqa}
D.~G. Figueroa, J.~Garcia-Bellido and F.~Torrenti, \emph{{Decay of the standard
  model Higgs field after inflation}},
  \href{https://doi.org/10.1103/PhysRevD.92.083511}{\emph{Phys. Rev.}
  {\bfseries D92} (2015) 083511},
  [\href{https://arxiv.org/abs/1504.04600}{{\ttfamily 1504.04600}}].

\bibitem{Enqvist:2015sua}
K.~Enqvist, S.~Nurmi, S.~Rusak and D.~Weir, \emph{{Lattice Calculation of the
  Decay of Primordial Higgs Condensate}},
  \href{https://doi.org/10.1088/1475-7516/2016/02/057}{\emph{JCAP} {\bfseries
  1602} (2016) 057}, [\href{https://arxiv.org/abs/1506.06895}{{\ttfamily
  1506.06895}}].

\bibitem{Figueroa:2017slm}
D.~G. Figueroa, A.~Rajantie and F.~Torrenti, \emph{{Higgs field-curvature
  coupling and postinflationary vacuum instability}},
  \href{https://doi.org/10.1103/PhysRevD.98.023532}{\emph{Phys. Rev. D}
  {\bfseries 98} (2018) 023532},
  [\href{https://arxiv.org/abs/1709.00398}{{\ttfamily 1709.00398}}].

\bibitem{Bezrukov:2014ipa}
F.~Bezrukov, J.~Rubio and M.~Shaposhnikov, \emph{{Living beyond the edge: Higgs
  inflation and vacuum metastability}},
  \href{https://doi.org/10.1103/PhysRevD.92.083512}{\emph{Phys. Rev. D}
  {\bfseries 92} (2015) 083512},
  [\href{https://arxiv.org/abs/1412.3811}{{\ttfamily 1412.3811}}].

\bibitem{Arnold:2002zm}
P.~B. Arnold, G.~D. Moore and L.~G. Yaffe, \emph{{Effective kinetic theory for
  high temperature gauge theories}},
  \href{https://doi.org/10.1088/1126-6708/2003/01/030}{\emph{JHEP} {\bfseries
  01} (2003) 030}, [\href{https://arxiv.org/abs/hep-ph/0209353}{{\ttfamily
  hep-ph/0209353}}].

\bibitem{Kurkela:2011ti}
A.~Kurkela and G.~D. Moore, \emph{{Thermalization in Weakly Coupled Nonabelian
  Plasmas}}, \href{https://doi.org/10.1007/JHEP12(2011)044}{\emph{JHEP}
  {\bfseries 12} (2011) 044},
  [\href{https://arxiv.org/abs/1107.5050}{{\ttfamily 1107.5050}}].

\bibitem{Kamada:2014qja}
A.~Kamada and M.~Yamada, \emph{{Gravitational waves as a probe of the SUSY
  scale}}, \href{https://doi.org/10.1103/PhysRevD.91.063529}{\emph{Phys. Rev.
  D} {\bfseries 91} (2015) 063529},
  [\href{https://arxiv.org/abs/1407.2882}{{\ttfamily 1407.2882}}].

\bibitem{Kamada:2015iga}
A.~Kamada and M.~Yamada, \emph{{Gravitational wave signals from short-lived
  topological defects in the MSSM}},
  \href{https://doi.org/10.1088/1475-7516/2015/10/021}{\emph{JCAP} {\bfseries
  1510} (2015) 021}, [\href{https://arxiv.org/abs/1505.01167}{{\ttfamily
  1505.01167}}].

\end{thebibliography}\endgroup

\end{document}